\documentstyle[aps]{revtex}
\begin{document}
\title{Second order gauge invariant gravitational perturbations
of a Kerr black hole}
\author{Manuela CAMPANELLI\thanks{
Electronic address: manuela@aei-potsdam.mpg.de} and Carlos O. LOUSTO%
\thanks{Electronic address: lousto@aei-potsdam.mpg.de}}
\address{Max-Planck-Institut f\"ur Gravitationsphysik,
Albert-Einstein-Institut,\\
Schlaatzweg 1, D-14473 Potsdam, Germany,\\
Institut f\"ur Astronomie und Astrophysik, Universit\"at
T\"ubingen,\\
D-72076 T\"ubingen, Germany,\\
and\\
Instituto de Astronom\'\i a y F\'\i sica del Espacio,\\
Casilla de Correo 67, Sucursal 28, 
(1428) Buenos Aires, Argentina}
\date{\today }
\maketitle

\begin{abstract}
We investigate higher than the first order gravitational perturbations in
the Newman-Penrose formalism. Equations for the Weyl scalar $\psi_4,$
representing outgoing gravitational radiation, can be uncoupled into a
single wave equation to any perturbative order. For second order
perturbations about a Kerr black hole, we prove the existence of a first and
second order gauge (coordinates) and tetrad invariant waveform, $\psi_I$, by
explicit construction. This waveform is formed by the second order piece of
$\psi_4$ plus a term, quadratic in first order perturbations, chosen to make 
$\psi_I$ totally invariant and to have the appropriate behavior in an
asymptotically flat gauge. $\psi_I$ fulfills a single wave equation of the
form ${\cal T}\psi_I=S,$ where ${\cal T}$ is the same wave operator as for
first order perturbations and $S$ is a source term build up out of (known to
this level) first order perturbations. We discuss the issues of imposition of
initial data to this equation, computation of the energy and momentum
radiated and wave extraction for direct comparison with full numerical
approaches to solve Einstein equations.
\end{abstract}

\section{Motivations and overview}

The prediction of accurate waveforms generated during the final orbital
stage of binary black holes has become a worldwide research topic in general
relativity during this decade. The main reason is that these catastrophic
astrophysical events, considered one of the strongest sources of
gravitational radiation in the universe, are potentially observable by LIGO,
VIRGO and other interferometric detectors.
For its strong nonlinear features this black hole merger problem is only
fully tractable by direct numerical integration (with supercomputers) of
Einstein equations. Several difficulties remain to be solved in this
approach such as the presence of early
instabilities in the codes for numerical evolution of 
Einstein theory \cite{S98}, and finding a new prescription for astrophysically
realistic initial data representing orbiting black holes \cite{LP97b,LP98}.
Meanwhile, perturbation theory has shown not only to be the main approximation
scheme for computation of gravitational radiation, but also a useful
tool to provide benchmarks for full numerical simulations. From the
theoretical point of view perhaps the more relevant contribution
during the nineties in perturbative theory has been the ``close limit
approximation'' \cite{PP94}. It considers the final merger state of two
black holes as described by a {\it single} perturbed one. This idea was
applied to the head-on collision of two black holes and the emitted
gravitational radiation was computed by means of the techniques used in
first order perturbation theory
around a Schwarzschild black hole. When the results of this computation have
been compared with those of the full numerical integration of Einstein
equations
the agreement was so good that it was disturbing \cite{APPSS95}. This
encouraged the significant effort invested into the development of a second
order Zerilli formalism of metric perturbations about the Schwarzschild
background. The method was successfully implemented with particular emphasis
on the comparison with the fully numerically generated results. In the case
of two initially stationary black holes (Misner data) the agreement of the
results is striking \cite{GNPP96}. Second order perturbation theory confirmed
the success of the close limit approximation with an impressive agreement in
both waveforms and energy radiated against the full numerical simulations.
There has been a tantamount success in the extension of these studies to the
case of initially moving towards each other black holes \cite{NGPP98}, and
for slowly rotating ones \cite{GNPP98} (See Ref. \cite{P98} for a
comprehensive review).

All the above close limit computations are based on the Zerilli \cite{Z70}
approach to metric perturbations of a Schwarzschild, i.e. nonrotating, black
hole. This method uses the Regge-Wheeler \cite{RW57} decomposition of the
metric perturbations into multipoles (tensor harmonics). Einstein equations
in the Regge-Wheeler gauge reduce to two single wave equations for the even
and odd parity modes of the gravitational perturbations. There is, however,
the strong belief that binary black holes in a realistic astrophysical
scenario merge together into a single, highly rotating, black hole. There is
also concrete observational evidence of accreting black holes \cite{chinos}
that places the rotation parameter as high as $a/M\simeq $ $0.95$. Finally,
highly rotating black holes provide a new scenario to compare perturbative
theory with full numerical integrations of Einstein equations.

The Regge-Wheeler-Zerilli techniques cannot be extended to study 
perturbations on a Kerr black hole background (see Ref.\ \cite{GNPP98}
for the slowly rotating case). In this case
there is not a multipole decomposition of metric perturbations
(in the time domain) and Einstein equations cannot be uncoupled into wave
equations. A reformulation of the gravitational field equations
due to Newman and Penrose \cite{NP62}, based on the Einstein equations and
Bianchi identities projected along a null tetrad, 
allowed Teukolsky\cite{T72} to write down a single master wave
equation for the perturbations of the Kerr metric in terms of the Weyl
scalars $\psi_4$ or $\psi_0$.
This formulation has several advantages: i) It is a first order gauge
invariant description. ii) It does not rely on any frequency or
multipole decomposition. iii) The Weyl scalars are objects defined in
the full nonlinear theory and a one
parameter perturbative expansion of it was proved to provide a reliable
account of the problem\cite{DS90}. In addition, the Newman-Penrose formulation
constitutes a simpler and more elegant framework to organize higher
order perturbation schemes as we will see in the next section.

Since the seventies the Teukolsky equation for the first order perturbations
around a rotating black hole has been Fourier transformed and integrated in
the frequency domain for a variety of situations where initial data played
no role (see Ref. \cite{FN89} for a review). Very recently it was
proved\cite{P97,CL97} that
nothing is intrinsically wrong with the Teukolsky equation when sources
extend to infinity and that a regularization method produces sensible results.
In order to incorporate initial
data and have a notable computational efficiency, concrete
progress has been made recently to complete a computational
framework that allows to
integrate the Teukolsky equation in the {\it time domain}: First, an
evolution code for integration of the Teukolsky wave equation is now
available\cite{KLPA97} and successfully tested\cite{CKL98}. Second, non
conformally flat Cauchy
data, compatible with Boyer-Lindquist slices of the Kerr geometry, began to
be studied with a Kerr-Schild\cite{BIMW97,MHS98} or an axially symmetric
\cite{BP98,KP98} ansatz. Finally, an
expression connecting $\psi_4$ to only Cauchy data has been worked out
explicitly\cite{CL98I,CKL98,CLBKP98}.

Assuming that we can solve for the first order perturbations problem, we
decided to go one step forward in setting the formalism for the second order
perturbations. As motivations for this work we can cite the spectacular
results presented in Ref.\ \cite{GNPP96} for the head-on collision and the
hope to obtain similar agreement for the orbital binary black hole case
in the close limit. Second order perturbations of the Kerr metric
may even play a
more important role in this case since we expect the perturbative
parameter to be linear in the separation of the holes\cite{KP98b}
while in the
head on case it is quadratic in the separation\cite{AP96b}. The
nonrotating limit of our approach will also provide an independent test and
clarify some aspects of Ref.\ \cite{GNPP96} results. High precision
comparison with full numerical integration of Einstein equations using
perturbative theory as benchmarks is also one of the main
goals in this program as well as a the development of a tool to explore
a complementary region of the parameter space to that reachable by full
numerical methods. 
An important application of second order perturbations is to provide error
bars. It is well known that linearized perturbation theory does not provide,
in itself, any indication on how good the perturbative approximation is. In
fact, it is in general very difficult to estimate the errors involved in
replacing an exact solution of the full Einstein equations with an
approximate (perturbative) solution, i.e., to determine how small a
perturbative parameter $\varepsilon $ must be in order that the approximate
solution have sufficient accuracy. Moreover, first order perturbation theory
can be very sensitive to the choice of parametrization, i. e. different
choices of the perturbative parameter can affect the accuracy of the
linearized approximation \cite{AP96a}. The only reliable procedure to resolve
the error and/or parameter arbitrariness is to carry out computations of the
radiated waveforms and energy to second order in the expansion parameter.
The ratio of second order corrections to the linear order results
constitutes the only direct and systematically independent measure of the
goodness of the perturbation results.

In the next section we extend to second (and higher) order the Teukolsky
derivation of the equation that describes
first order perturbations about a Kerr hole.
To do so we consider the Newman-Penrose \cite{NP62} formulation of the
Bianchi identities and Einstein equations, make a perturbative expansion
of it, and decouple the equation that
describes the evolution of second (and higher) order perturbations.
This equation takes the following form 
\begin{equation}
\widehat{{\cal T}}\psi^{(2)}=S[\psi ^{(1)},\partial _t\psi ^{(1)}],
\label{uno}
\end{equation}
where $\psi\dot=(\rho^{(0)})^{-4}\psi_4$, 
$\widehat{{\cal T}}$ is the same (zeroth order) wave operator that
applies to first order perturbations (see Eq.\ (\ref{master})) and $S$ is a
source term quadratic in the first order perturbations
(see Eqs.\ (\ref{fuente})-(\ref{dos})).

In Section III.A we describe how to compute the source, appearing in
Eq.\ (\ref{uno}), in terms of solutions of the wave equations for
$\psi_4^{(1)}$ or $\psi_0^{(1)}$ only, which are the objects we
directly obtain from the integration of the first order Teukolsky equation.
Sec.\ III.B discusses the issue of building up $\psi_{4}^{(2)}$ and 
$\partial_t\psi_{4}^{(2)}$ out of initial data (that we assume are given
to first and second order).
In section III.C we recall the equations for the computation of the
second order total radiated energy and momentum.

Higher than first order calculations are always characterized by an
extraordinary complexity and a number of subtle, potentially confusing,
gauge issues mainly due to the fact that a general second order gauge
invariant formulation is not yet at hand in the literature. In general,
gauge invariant quantities have an inherent physical meaning and they
automatically lead to the simpler and direct interpretation of the results.
In the Newman-Penrose formalism one has not only to look at gauge
invariance (i. e. invariance under infinitesimal coordinates
transformations), but also at invariance under tetrad rotations (see
Sections IV.A and IV.B). More specifically, the problem here is that the
waveform $\psi_4^{(2)}$ in Eq.\ (\ref{uno})) is neither first order
coordinate gauge invariant nor tetrad invariant. The question that arises
therefore is whether $\psi _4^{(2)}$ can be unambiguously compared with,
for instance, full numerical computations of the covariant $\psi_4^{Num}$.
To handle this problem we build up a coordinate and tetrad
invariant quantity up to second order, $\psi_{I}^{(2)}$, which
has the property of reducing to the linear part
(in the second order perturbations of the metric)
of $\psi_4^{(2)}$ in an asymptotically flat gauge at the ``radiation
zone'', far from the sources. This property ensures us direct
comparison with $\psi_4^{Num}$ by constructing
$\psi_4^{(1)}+\psi_{4\ I}^{(2)}$. In Sections
IV.A.-IV.C. we give an explicit and general prescription for the
construction of second order gauge and tetrad invariant objects representing
outgoing radiation. To do so we impose the waveform $\psi_{I}^{(2)}$
to be invariant under a ``combined'' transformation of both the
coordinates and the tetrad frame to first and second order.
The resulting second order invariant waveform can then be built
up out of the original $\psi_4^{(2)}$ plus corrections (quadratic in
the first order quantities) that cancel out the gauge and tetrad
dependence of $\psi_4^{(2)}$.
Finally, in Sec. V, along with a short summary, we
discuss the astrophysical and numerical applications of our result.
We end the paper with three
appendices: Appendix A refers Sec. III.A and contains explicit formulas to
compute the first order perturbative Newman-Penrose quantities (Weyl
scalars, spin coefficients and perturbed tetrad) in terms of the first order
metric perturbations needed to build up the source term in the wave equation
for $\psi _{{\rm I}}^{(2)}$. Appendix B refers to Sec.\ III.B and contains
formulas to compute the second order spin coefficients in terms of the
second order metric perturbations and product of first order perturbations
needed, for instance, to build up $\psi_{I}^{(2)}$ in terms of initial data.
Finally, in Appendix C we explicitly give the expressions to build up
the gauge invariant waveform holding in the Schwarzschild limit case, i.e.
for $a=0.$

Notation: In this paper we use Refs.\ \cite{T73,NP62} conventions. Background
quantities carry the (0) superindex if needed for clarity and are all
explicitly given in the cited references, while superindices (1) and (2)
mean pieces of {\it exclusively} first and second order respectively,
for instance, we expand $\psi=\psi^{(0)}+\psi^{(1)}+\psi^{(2)}+...$

\section{Decoupled equations for higher order gravitational perturbations}

Let us consider the following two of the eight complex Bianchi
identities written in the Newman-Penrose formalism (projected along a
complex null tetrad) \cite{C83}, Ch. 1.8 (see also the Appendix A) 
\begin{eqnarray}
&&\ \ \left( D+4\epsilon -\rho \right) \psi _4-\left( \overline{\delta }%
+4\pi +2\alpha \right) \psi _3+3\lambda \psi _2  \label{bianchi1} \\
\ &=&4\pi [\left( \overline{\delta }-2\overline{\tau }+2\alpha \right) T_{n%
\overline{m}}-\left( \Delta +2\gamma -2\overline{\gamma }+\overline{\mu }%
\right) T_{\overline{m}\overline{m}}\ -\lambda \left( T_{nl}+T_{m\overline{m}%
}\right) +\overline{\sigma }T_{nn}+\nu T_{l\overline{m}}],  \nonumber \\
&&  \nonumber \\
&&\ \ \left( \delta +4\beta -\tau \right) \psi _4-\left( \Delta +4\mu
+2\gamma \right) \psi _3+3\nu \psi _2  \label{bianchi2} \\
\ &=&4\pi [\left( \overline{\delta }-\overline{\tau }+2\overline{\beta }%
+2\alpha \right) T_{nn}-\left( \Delta +2\gamma +2\overline{\mu }\right) T_{n%
\overline{m}}\ \ +\nu \left( T_{nl}+T_{m\overline{m}}\right) +\overline{\nu }%
T_{\overline{m}\overline{m}}-\lambda T_{nm}]  \nonumber
\end{eqnarray}
and the following one out of the eighteen complex Ricci identities \cite{C83}
\begin{equation}
\left( \Delta +\mu +\overline{\mu }+3\gamma -\overline{\gamma }\right)
\lambda -\left( \overline{\delta }+3\alpha +\overline{\beta }+\pi -\overline{%
\tau }\right) \nu +\psi _4=0.  \label{ricci1}
\end{equation}
Here $D=l_{\ }^{\mu}\partial_\mu,
\Delta=n_{\ }^{\mu}\partial_\mu,
\delta=m_{\ }^{\mu}\partial_\mu$.

In what follows it is convenient to define the operators 
\begin{equation}
\overline{d}_3\dot =\left( \overline{\delta }+3\alpha +\overline{\beta }%
+4\pi -\overline{\tau }\right) ,~~\overline{d}_4\dot =\left( \Delta +4\mu +%
\overline{\mu }+3\gamma -\overline{\gamma }\right) .\label{d3d4}
\end{equation}

In order to find a decoupled equation for $\psi _4$ we operate
with\footnote{Here we use operators defined on the background instead of
(\protect\ref{d3d4}) for the sake of simplicity.}
$\overline{d}_4^{(0)}$ on Eq. (\ref{bianchi1}), with $\overline{d}_3^{(0)}$
on Eq. (\ref{bianchi2}), and then subtract to obtain 
\begin{eqnarray}
&&\left[ \overline{d}_4^{(0)}\left( D+4\epsilon -\rho \right) -\overline{d}%
_3^{(0)}\left( \delta +4\beta -\tau \right) \right] \psi _4  \nonumber \\
&&+\left[ \overline{d}_3^{(0)}\left( \Delta +4\mu +2\gamma \right) -%
\overline{d}_4^{(0)}\left( \overline{\delta }+4\pi +2\alpha \right) \right]
\psi _3  \nonumber \\
&&-3\left[ \overline{d}_3^{(0)}\nu -\overline{d}_4^{(0)}\lambda \right] \psi
_2  \nonumber \\
&=&T[\text{matter}],  \label{teuk0}
\end{eqnarray}
where $T[$matter$]$ is defined in Eq.\ (\ref{materia}) below.

In the above equation $\psi _4,\psi _3,\nu $ and $\lambda $ vanish on the
background, i.e. on the Kerr geometry, but so far this equation is exact, no
perturbative expansion has been made yet. Let us now think how to use Eq.
(\ref{teuk0}) in a perturbative scheme. In this context, the superindex $(p)$
appearing in the formulae below stands for a sum over all perturbative
orders from $p=1$ up to $p=n-1$ (i. e. $\sum_{p=1}^{n-1}$) where $n=1,2,...$
is an arbitrary order we want to study.

To fix ideas let us first discuss second order perturbations, $n=2$. The
procedure for higher order perturbations will be clearly analogous. We want
to have an uncoupled equation for $\psi_4^{(2)}$. Since $\psi_4^{(0)}=0$,
the operator in the first bracket on the left hand side of Eq.\ (\ref{teuk0})
is needed to zeroth plus first order. The zeroth order acts on $\psi_4^{(2)}$
and generates the same wave operator as for the first order
perturbations.
The first order operator in the first bracket on the left hand side of Eq.\ 
(\ref{teuk0}) acts on $\psi _4^{(1)}$ and its result can be considered as
generating an additional source term since it is supposed we have already
solved for the first order perturbation problem previously. The second
bracket on the left hand side of Eq. (\ref{teuk0}) can be considered as a
pure source term as well since its zeroth order vanishes 
\[
\overline{d}_3^{(0)}\left( \Delta +4\mu +2\gamma \right) ^{(0)}-\overline{d}%
_4^{(0)}\left( \overline{\delta }+4\pi +2\alpha \right) ^{(0)}=0, 
\]
(see Ref. \cite{T73} for an analogous proof) and then we have to consider
$\psi _3^{(1)}$, i.e. only to first perturbative order
(in general, to all lower perturbative orders than the one
considered). The last bracket on the left hand side of Eq. (\ref{teuk0})
includes terms depending on $\nu ^{(2)}$ and $\lambda ^{(2)}$ since $\psi
_2^{(0)}(=-M/(r-ia\cos \vartheta )^3)$ , is non vanishing. To get rid of these
second order spin coefficients we use Eq. (\ref{ricci1}) multiplied to the
left by $\psi _2^{(0)}$%
\[
\left[ \overline{d}_3^{(0)}\nu ^{(n)}-\overline{d}_4^{(0)}\lambda
^{(n)}\right] \psi _2^{(0)}=\psi _2^{(0)}\sum_{p=1}^{n-1}
\left[ \left( \overline{d}_3-3\pi
\right) ^{(n-p)}\nu ^{(p)}-\left( \overline{d}_4-3\mu \right)
^{(n-p)}\lambda ^{(p)}\right] +\psi _2^{(0)}\psi _4^{(n)}, 
\]
where we have made use of the pure zeroth order relations $\Delta ^{(0)}\psi
_2^{(0)}=-3\mu ^{(0)}\psi _2^{(0)}$ and $\overline{\delta }^{(0)}\psi
_2^{(0)}=-3\pi ^{(0)}\psi _2^{(0)}$ coming from the Bianchi identities. The
above result allow us again to write the terms depending on $\psi _2$ as
source terms.

We finally obtain the equation that describes the $n$-th order
perturbations 
\begin{equation}
\left\{ \overline{d}_4^{(0)}\left( D+4\epsilon -\rho \right)^{(0)}
-\overline{d}_3^{(0)}\left( \delta +4\beta -\tau \right)^{(0)}-3\psi_2^{(0)}
\right\}\psi _4^{(n)}={\cal S}%
_4[\psi ^{(n-p)},\partial _t\psi ^{(n-p)}]+T[\text{matter}],  \label{unodos}
\end{equation}
where 
\begin{equation}
\psi _4^{(n)}\doteq -(C_{\alpha \beta \gamma \delta }n^\alpha \overline{m}%
^\beta n^\gamma \overline{m}^\delta )^{(n)}
\end{equation}
and the source terms are (where brackets represent operators) 
\begin{eqnarray}
{\cal S}_4 &=&\sum_{p=1}^{n-1}\Bigg\{
\ \left[ \overline{d}_3^{(0)}\left( \delta +4\beta -\tau
\right) ^{(n-p)}-\overline{d}_4^{(0)}\left( D+4\epsilon -\rho \right)
^{(n-p)}\right] \psi _4^{(p)}  \nonumber \\
&&\ \ -\left[ \overline{d}_3^{(0)}\left( \Delta +4\mu +2\gamma \right)
^{(n-p)}-\overline{d}_4^{(0)}\left( \overline{\delta }+4\pi +2\alpha \right)
^{(n-p)}\right] \psi _3^{(p)}  \nonumber \\
&&\ \ +3\left[ \overline{d}_3^{(0)}\nu ^{(n-p)}-\overline{d}_4^{(0)}\lambda
^{(n-p)}\right] \psi _2^{(p)}  \label{fuente} \\
&&\ -3\psi _2^{(0)}\left[ \left( \overline{d}_3-3\pi \right) ^{(n-p)}\nu
^{(p)}-\left( \overline{d}_4-3\mu \right) ^{(n-p)}\lambda ^{(p)}\right]
\Bigg\} , 
\nonumber
\end{eqnarray}
and 
\begin{eqnarray}
T[\text{matter}] &=&\sum_{p=1}^{n-1}\Bigg\{
\overline{d}_4^{(0)}\left[ \left( \overline{\delta }-2%
\overline{\tau }+2\alpha \right) ^{(n-p)}T_{n\overline{m}}^{(p)}-\left(
\Delta +2\gamma -2\overline{\gamma }+\overline{\mu }\right) ^{(n-p)}T_{%
\overline{m}\overline{m}}^{(p)}\right]  \nonumber \\
&&+\overline{d}_3^{(0)}\left[ \left( \Delta +2\gamma +2\overline{\mu }%
\right) ^{(n-p)}T_{n\overline{m}}^{(p)}-\left( \overline{\delta }-\overline{%
\tau }+2\overline{\beta }+2\alpha \right) ^{(n-p)}T_{nn}^{(p)}\right]\Bigg\},
\label{materia}
\end{eqnarray}
where $T_{n\overline{m}}^{(p)}=(T_{\mu \nu }n^\mu \overline{m}^\nu
)^{(p)},T_{\overline{m}\overline{m}}^{(p)}=(T_{\mu \nu }\overline{m}^\mu 
\overline{m}^\nu )^{(p)}$ and $T_{nn}^{(p)}=(T_{\mu \nu }n^\mu n^\nu
)^{(p)}. $ Note that in our formalism we have taken into account matter
terms in order to be used in future computations including an orbiting
particle or an accretion disk around a Kerr hole. By summing up over
all $n-$orders in Eq.\ (\ref{unodos}) one should be able to recover
solutions to the full Einstein Equations.

Note also that if one wants to act on
$\psi ^{(n)}\dot =\rho ^{-4}\psi _4^{(n)}$
rather than $\psi _4^{(n)}$ one should rescale all the terms (including
the source) in Eq. (\ref{unodos}) by a factor $2\rho ^{-4}\Sigma $. After
this rescaling, Eq. (\ref{unodos}) takes the following familiar form 
\begin{equation}
\widehat{{\cal T}}\psi ^{(n)}=2\rho ^{-4}\Sigma \{{\cal S}_4[\psi
^{(n-p)},\partial _t\psi ^{(n-p)}]+T[\text{matter}]\}.  \label{dos}
\end{equation}
In Ref.\ \cite{T73} the wave operator was
transformed to act on the field $\psi^{(1)}\dot=(\rho^{(0)})^{-4}\psi_4^{(1)}$
rather than $\psi_4^{(1)}$ (in order to achieve separability of the variables
in the frequency domain) and takes the following form, in
Boyer-Lindquist coordinates $(t,r,\vartheta ,\varphi )$ and Kinnersley
tetrad
\begin{eqnarray}
&&\widehat{{\cal T}}=\left[ \frac{(r^2+a^2)^2}\Delta -a^2\sin ^2\vartheta
\right] \partial _{tt}+\frac{4Mar}\Delta \partial _{t\varphi }-4\left[
r+ia\cos \vartheta -\frac{M(r^2-a^2)}\Delta \right] \partial _t  \nonumber \\
&&-\,\Delta ^2\partial _r\left( \Delta ^{-1}\partial _r\right) -\frac 1{\sin
\vartheta }\partial _\vartheta \left( \sin \vartheta \partial _\vartheta
\right) -\left[ \frac 1{\sin ^2\vartheta }-\frac{a^2}\Delta \right] \partial
_{\varphi \varphi }  \label{master} \\
&&+\,4\left[ \frac{a(r-M)}\Delta +\frac{i\cos \vartheta }{\sin ^2\vartheta }%
\right] \partial _\varphi + \left( 4\cot ^2\vartheta +2\right) ,  \nonumber
\end{eqnarray}
where $M$ is the mass of the black hole, $a$ its angular momentum per unit
mass, $\Sigma \equiv r^2+a^2\cos ^2\vartheta $, and 
$\Delta \equiv r^2-2Mr+a^2$.
Note that if one wants to act on $\psi^{(2)}\dot=(\rho^{(0)})^{-4}\psi_4^{(2)}$
rather than $\psi_4^{(2)}$ in Eq. (\ref{teuk0}) then one should
consistently rescale all the terms (including the source) by a factor 
$2(\rho^{(0)})^{-4}\Sigma$ (see Eq. (\ref{dos})).

It is easy to show that a similar equation to (\ref{dos}) can be
obtained for the Weyl scalar field $\psi _0$, upon exchange of the tetrad
vectors $l\leftrightarrow n$ and $\overline{m}\leftrightarrow m$. In this paper
we will explicitly work with $\psi _4$ since it
directly represents outgoing gravitational radiation.
Since at every level of the hierarchy of perturbations we have the zeroth
order wave operator acting on $\psi_4^{(n)}$ we could always use the method
of full separations of variables. In this paper, however,
we will not proceed so because we want our equations to be suitable
for evolution in the time domain from given Cauchy data.

\section{practical issues}

\subsection{Gauge choice and computation of the source}

As we will show explicitly in the next Section, $\psi_4$ is neither invariant
under first order coordinates transformations nor second order tetrad
rotations.
Thus, in order to integrate Eq.\ (\ref{dos}), one would have to evolve
$\psi$ in a fixed gauge (and tetrad) and then compute physical quantities,
like radiated energy and waveform, in an asymptotically flat gauge.
This sort of approach was followed in Ref.\ \cite{GNPP98} to study
second order perturbations of a Schwarzschild black hole in the Regge-Wheeler
gauge which is a unique gauge that allows to invert expressions in terms
of generic perturbations and thus recover the gauge invariance.
There is not a generalization of the Regge-Wheeler gauge when studying
perturbations of a Kerr hole, essentially because one cannot perform a
simple multipole decomposition of the metric. Instead, Chrzanowski
\cite{C75} found two convenient gauges that allowed him to invert the
metric perturbations in terms of potentials $\Psi_{IRG}$ or $\Psi_{ORG}$
satisfying the same wave equations as the Weyl scalars $\rho^{-4}\psi_4$
or $\psi_0$ respectively.

In the {\it ingoing radiation} gauge (IRG)
\begin{equation}
h^{(1)}_{ll}=0=h^{(1)}_{ln}=0=h^{(1)}_{lm}=0=h^{(1)}_{l\overline{m}}=0=
h^{(1)}_{m\overline{m}} ,  \label{incalibre}
\end{equation}
the homogeneous (for vacuum) metric components can be written,
in the time domain, in terms of solutions to the wave equation for
$\rho^{-4}\psi_4^{(1)}$ only, as follows 
\begin{eqnarray}
(h^{(1)}_{\mu\nu})_{\ IRG}&=&2Re\Bigg[\left\{-l_\mu l_\nu(\delta +\overline{%
\alpha }+3\beta -\tau) (\delta +4\beta+3\tau)-m_\mu
m_\nu(D-\rho)(D+3\rho)\right.  \nonumber \\
&&\left.+l_{(\mu}m_{\nu)} [(D+\overline{\rho}-\rho)(\delta +4\beta+3\tau)+
(\delta-\overline{\alpha }+3\beta-\overline{\pi }-\tau)(D+3\rho)]
\right\}(\Psi_{IRG})\Bigg]\label{inmetrica}
\end{eqnarray}
where $Re$ stands for the real part of the whole object to ensure
that the metric be real\cite{W78,CK79} and we made the $\epsilon=0$ choice.
Note that in this gauge the metric potential has the property to be
transverse $(h^{(1)}_{\mu\nu}l^\mu=0)$ and traceless $(h^{(1)~\mu}_\mu=0)$
at the future horizon and past infinity. This is thus a suitable gauge to
study gravitational radiation effects near the event horizon.

The complementary (adjoint) gauge to the ingoing radiation gauge is the
{\it outgoing radiation} gauge (ORG), 
\begin{equation}
h^{(1)}_{nn}=0=h^{(1)}_{ln}=0=h^{(1)}_{nm}=0=h^{(1)}_{n\overline{m}}=0=
h^{(1)}_{m\overline{m}} .  \label{outcalibre}
\end{equation}
where the metric potential has now the property to be transverse 
$(h^{(1)}_{\mu\nu}n^\mu=0)$ and traceless $(h^{(1)~\mu}_\mu=0)$ at the
past horizon and future infinity. It is then an example of a suitable
asymptotically flat
gauge to directly compute radiated energy and momenta at infinity
(see Sec.\ III.C).
In this gauge, the homogeneous metric components can be written in terms of
solutions to the wave equation for $\psi_0^{(1)}$, as
\begin{eqnarray}
(h^{(1)}_{\mu\nu})_{\ ORG}&=&2Re\Bigg[
\rho^{-4} \left\{-n_\mu n_\nu (\overline%
\delta -3\alpha -\overline\beta +5\pi) (\overline\delta -4\alpha+\pi)-%
\overline{m}_\mu \overline{m}_\nu (\Delta+5\mu-3\gamma+\overline\gamma%
)(\Delta+\mu-4\gamma)\right.  \nonumber \\
&&\left.+n_{(\mu}\overline{m}_{\nu)} [(\overline\delta -3\alpha+\overline%
\beta+5\pi+\overline\tau) (\Delta+\mu-4\gamma)+(\Delta+5\mu-\overline\mu%
-3\gamma-\overline\gamma) (\overline\delta-4\alpha +\pi)]
\right\}(\Psi_{ORG})\Bigg]\label{outmetrica}
\end{eqnarray}

Note that Eqs.\ (\ref{incalibre})
(or \ (\ref{outcalibre})) are {\it four} conditions on the {\it real}
part of the metric.
Although \ (\ref{incalibre})) (or \ (\ref{outmetrica})) do not
fix completely the gauge freedom, Chrzanowski metric choice given in
Eq. \ (\ref{inmetrica})) (or Eq. \ (\ref{outmetrica})), being a specific
choice between all the possible solutions satisfying those conditions,
{\it does uniquely fix} all of the extra freedom.
\bigskip

The potentials $\Psi_{IRG}$ and $\Psi_{ORG}$ fulfil the Teukolsky equation
for $\rho^{-4}\psi_4$ and $\psi_0$ respectively. To determine them we can
invert expressions Eqs.\ (\ref{incalibre}) or (\ref{outcalibre}) and its
time derivatives at the initial Cauchy surface to relate the potential to our
first order initial data. Alternatively, one can use the relations of these
potentials to gauge invariant objects like $\psi_0$ or $\rho^{-4}\psi_4$.
For instance, in the IRG we can take the relation $\psi_0=DDDD\Psi_{IRG}$
(See Eq.\ (5.28) of Ref.\ \cite{CK79}) or in the ORG the adjoint
relation $\psi_4=\Delta\Delta\Delta\Delta\Psi_{ORG}$. Here we lower the
order of the time derivatives of $\Psi$ to first order ones by repeated
use of the Teukolsky equation potentials fulfill (See, for instance,
Eq.\ (5.20) of Ref.\ \cite{CK79}). Since one can always make a mode
decomposition of the $\varphi$ dependence one ends up with a set of 
potential equations for $\Psi(r,\theta)$ and $\partial_t\Psi(r,\theta)$
at the initial time. Boundary conditions are chosen such that we obtain
bounded solutions. The numerical integration of these equations is
left for a forthcoming paper\cite{CKL99}. These solutions give us the
initial data to integrate the wave equations and then build up metric
perturbations form Eqs.\ (\ref{incalibre}) or (\ref{outcalibre}).
The imposition of initial data to $\psi_4$ and $\psi_0$ is discussed in the
next subsection.

Finally, in order to integrate Eq.\ (\ref{dos}) we assumed the knowledge of
the source term (\ref{fuente}) since it depends only on first order
perturbations. In practice, one solves the Teukolsky equation for
$\psi_4^{(1)}$ (and / or $\psi_0^{(1)})$ and builds up metric perturbations.
It then remains the task of
writing all first order Newman-Penrose quantities in terms of $h_{\mu\nu}$. 
This is not a trivial task, so we give all the equations relating the
Newman-Penrose fields to the metric perturbations in appendix A.

\subsection{Imposition of initial data}

To start the evolution one has to be able to impose initial data to the
second order invariant waveform.
We first note that, from its definition, we can write
\begin{equation}
\psi _4^{(2)}=-C^{(2)}_{n\overline{m}n\overline{m}}+\frac14
h^{(1)}_{nn}h^{(1)}_{\overline{m}
\overline{m}}\left(\psi_2^{(0)}+\overline{\psi}_2^{(0)}\right)
-2\left(h^{(1)}_{ln}-\frac12 h^{(1)}_{m\overline{m}}\right)\psi_4^{(1)}
-2h^{(1)}_{n\overline{m}}\psi_3^{(1)}.
\label{psi4segundo}
\end{equation}
For the sake of definiteness we have used here Eq.\ (\ref{A1}) choice
of the first
order tetrad, but it is clear that the above expression can be written
in a generic tetrad. Besides, since we are going to build up the invariant
$\psi_{I}^{(2)}$, any choice of the tetrad (and the gauge) leads to the
same, correct, result.

In Ref.\ \cite{CLBKP98} we have completely expressed $\psi_4^{(1)}$ (and its
time derivative) in terms of hypersurface data only.
The expression\footnote{Note that the factor 8 appearing in front of the
second bracket corrects an obvious misprint in
Ref.\ \protect{\cite{CLBKP98}}. This also applies to the Eq.\ (3.2) for 
$\partial_t\psi_4^{(1)}.$}
\begin{eqnarray}
C_{n\overline{m}n\overline{m}}
&=-&\left[\ {}^{(3)}{R}_{ijkl}+2K_{i[k}K_{l]j}\right]\hat n^i%
\hat{\overline{m}}^j\hat n^k\hat{\overline{m}}^l
+8N\left[ K_{j[k,l]}+\ {}^{(3)}{%
\Gamma }_{j[k}^pK_{l]p}\right]\hat n^{[0}\hat{\overline{m}}^{j]}
\hat n^k\hat{\overline{m}}^l
\nonumber\\
&&\ -4N^2\left[ \ {}^{(3)}{R}_{jl}-K_{jp}K_l^p+KK_{jl}-T_{jl}+\frac 
12Tg_{jl}\right]\hat n^{[0}\hat{\overline{m}}^{j]}\hat n^{[0}
\hat{\overline{m}}^{l]}
\end{eqnarray}
and its time derivative hold in general, to all order. Here
$N=(-g^{tt})^{-1/2}$, $N^i=N^2g^{ti}$, $\hat n^\mu=n^\mu+N^in^t$ and
$\hat{\overline{m}}^\mu=\overline{m}^\mu+N^i\overline{m}^t$. When we
expand the above relation to a given perturbative order $n$, the proof
given in Ref.\ \cite{CLBKP98} implies that $\psi_4$ and $\partial_t\psi_4$
will be independent on the lapse and shift of order $n$ (but will depend, of
course, on all lower perturbative orders of $N$ and $N^i$).

To express our second order object $\psi _{I}^{(2)}$ in
terms of the three-metric and the extrinsic curvature of the initial
hypersurface we will proceed as in Ref.\ \cite{CLBKP98} taking now into
account the additional terms, quadratic in the
first order perturbations. We then find
\begin{eqnarray}
C^{(2)}_{n\overline{m}n\overline{m}}
&=-&\left[\ {}^{(3)}{R}_{ijkl}+2K_{i[k}K_{l]j}\right]_{(2)}\hat n^i%
\hat{\overline{m}}^j\hat n^k\hat{\overline{m}}^l+8N_{(0)}\left[ K_{j[k,l]}
+\ {}^{(3)}{%
\Gamma }_{j[k}^pK_{l]p}\right] _{(2)}\hat n^{[0}\hat{\overline{m}}^{j]}
\hat n^k\hat{\overline{m}}^l
\nonumber\\
&&\ -4N_{(0)}^2\left[ \ {}^{(3)}{R}_{jl}-K_{jp}K_l^p+KK_{jl}-T_{jl}+\frac 
12Tg_{jl}\right] _{(2)}\hat n^{[0}\hat{\overline{m}}^{j]}\hat n^{[0}
\hat{\overline{m}}^{l]}
\nonumber\\
&&\ +8N_{(1)}\left[ K_{j[k,l]}+\ {}^{(3)}{%
\Gamma }_{j[k}^pK_{l]p}\right] _{(1)}\hat n^{[0}\hat{\overline{m}}^{j]}
\hat n^k\hat{\overline{m}}^l
\label{Cdos} \\
&&\ -8N_{(0)}N_{(1)}
\left[ \ {}^{(3)}{R}_{jl}-K_{jp}K_l^p+KK_{jl}-T_{jl}+\frac 
12Tg_{jl}\right]_{(1)}\hat n^{[0}\hat{\overline{m}}^{j]}\hat n^{[0}
\hat{\overline{m}}^{l]}\nonumber
\end{eqnarray}
Note that the first three
terms have the same structure as in the first order case [for terms linear in
$h^{(2)}_{ij}$ and $K^{(2)}_{ij}$.] There is not dependence on the 
second order lapse and shift, but $N_{(1)}$ and
the perturbative shift now explicitly appear.
To re-express them in terms of hypersurface data, we can make use of
Eq.\ (\ref{inmetrica}) and Appendix A expressions that relate all first order
quantities to $\Psi_{ORG}$, directly expressible in terms
of hypersurface data only as discussed before. And the same
technique allow us to build up the additional quadratic terms
occurring in $\psi_{I}^{(2)}$. Since the total $\psi_{I}^{(2)}$ was originally
invariant, its final expression is not affected by the use of the
a gauge choice (such as (\ref{inmetrica}) or (\ref{outmetrica}))
at an intermediate step.

For $\partial_t\psi _{I}^{(2)}$, the procedure is the same as before. We
note that terms linear in $h^{(2)}_{ij}$ and $K^{(2)}_{ij}$ will
have the same structure as in the first order case, so Eq.\ (3.2) of
Ref.\ \cite{CLBKP98} applies upon change of the subindex (1) by (2).
The additional terms, quadratic in the first order perturbations, can
be directly written in terms of $\partial_t\Psi_{IRG}$ by taking the
time derivative of Eq.\ (\ref{inmetrica}) and Appendix A expressions.

In Appendix B we give an independent derivation relating $\psi_4^{(2)}$
to the four-geometry. We split
\begin{equation}
\psi _4^{(2)}\doteq\psi _{4\text{L}}^{(2)}+\psi _{4\text{Q}}^{(2)},
\label{segundo}
\end{equation}
where the first term on the right hand side is linear in the second order
perturbations of the metric, i.e. $h_{\mu\nu}^{(2)}$ and is formally the
same as $\psi_4^{(1)}$ replacing $h_{\mu\nu}^{(2)}\to h_{\mu\nu}^{(1)}$.
The second term on the right hand side, i.e. $\psi_{4\text{Q}}^{(2)}$,
accounts for the quadratic part in first order metric perturbations.

\subsection{Radiated energy and momenta}

The energy and momenta radiated at infinity to second perturbative order
can be computed using the standard methods of linearized gravity 
(here $h_{\mu\nu}$ stands for $h_{\mu\nu}^{(1)}+h_{\mu\nu}^{(2)}+...$ 
defined in asymptotically flat coordinates at future null infinity).
For outgoing waves\cite{T73}
\begin{equation}
\lim_{r\to\infty}\psi_4=-\frac12(\partial^2_t h_{\hat{\vartheta}
\hat{\vartheta}}-i\partial^2_t h_{\hat\vartheta\hat\varphi}),
\end{equation}
the total radiated energy per unit time $(u=t-r)$
can thus be obtained from
the Landau-Lifschitz pseudo tensor as 
\begin{equation}
\frac{dE}{du}=\lim_{r\to\infty}\left\{ \frac{r^2}{4\pi}%
\int_{\Omega}d\Omega\left| \int_{-\infty}^{u}d\tilde{u}\ \psi_4(\tilde{u%
},r,\vartheta,\varphi) \right|^2\right\}, \quad
d\Omega=\sin\vartheta\ d\vartheta\ d\varphi,
 \label{energy}
\end{equation}
where we can consider $\psi_4=\psi_4^{(1)}+\psi_{4}^{(2)\ AF}+...$

Note that Eq.\ (\ref{energy}) can be equivalently obtained by directly
calculating the Bondi definition of the mass carried away by the
gravitational radiation by imposing asymptotically flat conditions to
the full Newman-Penrose quantities in a general vacuum spacetime.
In this way, one can also compute the total linear momentum radiated at
infinity per unit time along cartesian-like coordinates as\cite{NT80}
\begin{eqnarray}
&&\frac{dP_\mu}{du}=-\lim_{r\to\infty}\left\{ \frac{r^2}{4\pi}%
\int_{\Omega}d\Omega\ \tilde{l}_\mu\left| \int_{-\infty}^{u}d\tilde{u}\ 
\psi_4(\tilde{u},r,\vartheta,\varphi) \right|^2\right\},\label{momentum}\\
&&\tilde{l}_\mu=(1,-\sin\theta\cos\varphi,-\sin\theta\sin\varphi,-\cos\theta),
\nonumber
\end{eqnarray}
and the angular momentum carried away by the waves\cite{W80} can be obtained
from
\begin{equation}
\frac{dJ_z}{du}=-\lim_{r\to\infty}\left\{ \frac{r^2}{4\pi}\ Re\left[
\int_\Omega d\Omega
\left(\partial_\varphi\int_{-\infty}^{u}d\tilde{u}\
\psi_4(\tilde{u},r,\vartheta,\varphi) \right)
\left(\int_{-\infty}^{u}du^\prime\int_{-\infty}^{u^\prime}d\tilde{u}\ 
\overline{\psi}_4(\tilde{u},r,\vartheta,\varphi)\right)\right]
\right\}.  \label{angmomentum}
\end{equation}

One can directly compute the second order correction to the energy and
momentum radiated at ${\cal J}^+$
using $\psi_4^{(2)}$, provided one is working
(to first order) in an asymptotically flat gauge (for instance, the outgoing
radiation gauge). Eqs.\ (\ref{energy})--(\ref{angmomentum}), written in
terms of the full, nonlinear $\psi_4$, are covariant expressions, holding
in any asymptotically flat spacetime. To first perturbative order,
$\psi_4^{(1)}$ is directly gauge and tetrad invariant, so one can forget
that the above equations had been obtained in an asymptotically flat gauge
and think of them as gauge (and tetrad) invariant. We would like to have
the same nice property to second perturbative order, but $\psi_4^{(2)}$
is not invariant. One should then build up a gauge
(and tetrad) invariant
waveform $\psi_I^{(2)}$ that, in an asymptotically flat gauge coincides
with $\psi_4^{(2)\ AF}$. This will ensure us the direct use of
Eqs.\ (\ref{energy})--(\ref{angmomentum}) in terms of
our invariant object, i.e. $\psi_I^{(2)}$ given in Eq.\ (\ref{GIW}).

\section{Construction of the second order coordinate and tetrad invariant
waveform}

The general covariance (i.e. diffeomorphism invariance) of Einstein's theory
of gravity guarantees the complete freedom in the choice of the spacetime
coordinates (gauge) to describe physical phenomena. In the relativistic
theory of perturbations one always introduces two spacetimes, the physical
(perturbed) spacetime and an idealized (unperturbed) background. In this way
the perturbations can be viewed as fields propagating on the background.
Consequently, to compare any physical quantity in the perturbed spacetime
with the same quantity in the unperturbed spacetime it is necessary to
introduce a diffeomorphism about the pairwise identification points between
the two manifolds. The arbitrariness in the choice of this point
identification map introduces an additional freedom to the usual gauge
freedom of general relativity and is at the origin of the {\it gauge problem}
in perturbation theory \cite{BS98}. A convenient way to deal with this gauge
problem is to construct quantities which are invariant under a change of the
identification map of the perturbed spacetime while the background
coordinates are held fixed.

Invariance in the Newman-Penrose formalism has a more restrictive meaning
than in the standard (metric) perturbation theory, since the introduction of
a tetrad frame at every point of the spacetime now requires that any
physical perturbation must be invariant not only under infinitesimal gauge
transformations (GI), but also under infinitesimal rotations of the local
tetrad frame (TI). In this Section we briefly review the basic concepts of
(higher order) tetrad invariance and coordinate (gauge) invariance in the
framework of the Newman-Penrose formalism. We start with our second order
object $\psi _4^{\left( 2\right) }$ , which is neither invariant under first
order changes of the coordinates nor under second order tetrad rotations. We
then show how to build up a tetrad invariant object by adding to $\psi
_4^{\left( 2\right) }$ a conveniently chosen term, quadratic in the first
order perturbations. In this way the new object will be invariant under the
(6-parameter) tetrad rotations. The procedure for the construction of the
totally invariant object, i.e. also under coordinate choices (4-parameters)
is analogous, but algebraically more involved. The final
result is a general prescription for constructing totally invariant (I)
quantities directly related to the (outgoing) gravitational radiation.

\subsection{Tetrad invariance}

The 6-parameter group of homogeneous Lorentz transformations, which
preserves the tetrad orthogonality relations $l_\mu $ $n^\mu =-m_\mu 
\overline{m}^\mu =1$ (and all other scalar products zero), can be decomposed
into three Abelian subgroups:

\begin{itemize}
\item  Null rotation of type (I) which leaves the $l_\mu $ unchanged: 
\begin{eqnarray}
\widetilde{l_\mu }&\rightarrow& l_\mu \nonumber \\
\widetilde{n_\mu }&\rightarrow& n_\mu +\text{a}\overline{m}_\mu +
\overline{\text{a}}m_\mu +\text{a}\overline{\text{a}}l_\mu , \\
\widetilde{m_\mu }&\rightarrow& m_\mu +\text{a}l_\mu ;\nonumber
\end{eqnarray}

\item  Null rotation of type (II), which leaves the $n_\mu $ unchanged: 
\begin{eqnarray}
\widetilde{l_\mu }&\rightarrow& l_\mu +\text{b}\overline{m}_\mu +
\overline{\text{b}}m_\mu +\text{b}\overline{\text{b}}n_\mu , \nonumber\\
\widetilde{n_\mu }&\rightarrow& n_\mu , \\
\widetilde{m_\mu }&\rightarrow& m_\mu +\text{b}n_\mu ;\nonumber
\end{eqnarray}

\item  Boost and rotation of type (III): 
\begin{eqnarray}
\widetilde{l_\mu }&\rightarrow& Al_\mu , \nonumber\\
\widetilde{n_\mu }&\rightarrow& A^{-1}n_\mu , \\
\widetilde{m_\mu }&\rightarrow& \exp (i\theta )m_\mu ;\nonumber
\end{eqnarray}
\end{itemize}
where $(\text{a},\text{b})$ are two complex functions and $(A,\theta)$
two real functions
on the four dimensional manifold, hence the six arbitrary parameters. When
these functions are taken to be infinitesimally small the above
transformations can be expanded up to an arbitrary order and then applied
to any Newman-Penrose quantity.

Under a combined tetrad rotation of classes I, II, and III 
\begin{equation}
\widetilde{\psi_4^{(2)}}\rightarrow \psi_4^{(2)}+2[(A-1)-i\theta
]\psi_4^{(1)} +4\overline{\text{a}}\psi_3^{(1)}+
6\overline{\text{a}}^2\psi_2^{(0)}.
\label{psi42}
\end{equation}

The idea here is to supplement $\psi_4^{(2)}$ with additional terms that make
the whole object tetrad invariant. Since we have to add those ``correcting''
terms on both sides of the field equation (\ref{dos}), we will write them as
powers of first order perturbations so they can be added to the source term 
(\ref{fuente}). The first step towards constructing this quantity is to note
that $(l^r)^2(m^\vartheta )^2\psi _4/[(l^{r~(0)})^2(m^{\vartheta ~(0)})^2]$
is invariant under rotations of class III. The second
order piece of this combination of fields is 
\begin{equation}
\psi _4^{(2)}+2\left( \frac{l^{r~(1)}}{l^{r~(0)}}+\frac{m^{\vartheta ~(1)}}{%
m^{\vartheta ~(0)}}\right) \psi _4^{(1)}.  \label{III}
\end{equation}
Note that the second addend exactly compensates for the variation of class
III of $\psi _4^{(2)}$ (proportional to the parameters $A-1$ and $\theta $
in Eq. (\ref{psi42})). In addition one can easily check that the second term
in (\ref{III}) is also invariant under rotations of class I and II
with the Kinnersley choice\cite{T73} of the zeroth order tetrad
\begin{eqnarray}\label{t0}
   (l^{\mu})^{(0)} &=& \left( \frac{r^2+a^2}{\Delta},1,0,\frac{a}{\Delta} 
                  \right) \; , \nonumber\\
   (n^{\mu})^{(0)} &=& \frac{1}{2\;\!(r^2+a^2\cos^2\vartheta)} \,
                  \left( r^2+a^2,-\Delta,0,a \right) \; , \\
   (m^{\mu})^{(0)} &=& \frac{1}{\sqrt{2}(r+ia\cos\vartheta)}\,
                  \left( ia\sin\vartheta,0,1,i/\sin\vartheta \right)\; ,
\nonumber
\end{eqnarray}
since $l^{r~(1)}$, $m^{\vartheta ~(1)}$ and $\psi _4^{(1)}$ are
all invariant\footnote{It is clear that we can write the 
tetrad invariant object in terms of a generic zeroth order tetrad
by replacing in Eq.\ \protect{(\ref{QTI})}
$l^{r\ (1)}\to l^{r\ (1)}-m^r\overline{\psi}_1^{(1)}/(3
\overline{\psi}_2^{(0)})-
\overline{m}^r\psi_1^{(1)}/(3\psi_2^{(0)})$ and 
$m^{\vartheta\ (1)}\to m^{\vartheta\ (1)}-
l^\vartheta\psi_3^{(1)}/(3\psi_2^{(0)}).$ 
We take the background tetrad \protect{(\ref{t0})} for
the sake of simplicity.} under
rotations of class I and II. Still, the first term in (\ref{III}) varies
with respect to rotations of class I and II. To correct that we note that
under combined rotations I, II, and III 
\begin{eqnarray}
\widetilde{\psi _3^{(1)}}\rightarrow \psi _3^{(1)}+3\overline{\text{a}}
\psi _2^{(0)}.
\end{eqnarray}
This allows us to solve for $\overline{\text{a}}$ and replace it into the new
expression (its form suggested by the $\overline{\text{a}}$ dependence in the
transformation (\ref{psi42})) that supplement (\ref{III}). [Note that this
replacement is successful because $\psi _3$ vanishes to zeroth order.]
Thus, the object,
\begin{equation}
\psi _4^{(2)}+2\psi _4^{(1)}\left( \frac{
l^{r~(1)}}{l^{r~(0)}}+\frac{m^{\vartheta ~(1)}}{m^{\vartheta ~(0)}}\right)-
\frac 23\frac{\left( \psi _3^{(1)}\right) ^2}{\psi _2^{(0)}},
\label{QTI}
\end{equation}
is second order tetrad invariant.

While the above combination is tetrad invariant, one can see from the
general behavior of the Weyl scalars and spin coefficients in an
asymptotically flat gauge (see, for instance Sec. VII of Ref.\ \cite{NP62}),
that the quadratic term we added does not vanish relative to $\psi_4^{(2)}$
for large $r$, i.e. goes like ${\cal O}(1/r)$ as well. In order to have
the desired property that in the radiation zone the invariant object
approaches $\psi_4^{(2)\ AF}$ ($AF$ stands for an asymptotically flat gauge),
we will subtract to (\ref{QTI})
another quadratic part that both, cancels its added asymptotic behavior and
is tetrad (and gauge) invariant in order to preserve the gained invariance of
(\ref{QTI}).
Symbolically, if we call $Q$ the quadratic part we added to $\psi_4$
in Eq.\ (\ref{QTI}), we search for a
\begin{equation}
\psi_{4\ {\rm TI}}^{(2)}\dot=\psi_4^{(2)}+Q-Q_I^{AF}
\label{invariante}
\end{equation}
A practical way to build up $Q_I^{AF}$ is to use
relations (\ref{outmetrica}), i.e. the perturbed metric in the
outgoing radiation gauge, which is
an asymptotically flat gauge at infinity. In this gauge, we evaluate
the quadratic part $Q$ in (\ref{QTI}) and once, re-expressed all in terms 
of $\Psi_{ORG}$ via Eqs.\ (\ref{outmetrica}), we can forget that we
used the outgoing radiation gauge and see $Q_I^{ORG}$ as a tetrad and gauge
invariant object, since $\Psi_{ORG}$ is totally invariant.
In the outgoing radiation gauge $\psi_4^{(2)}$ (and $\psi_{4\ TI}^{(2)}$)
reduces to $\psi_{4\ L}^{(2)}$ as can be directly deduced from the
expressions given in Appendix B.

\subsection{Gauge invariance}

The meaning of gauge invariance under infinitesimal coordinate changes, to
an arbitrary order in the perturbations, was explicitly elucidated in
Ref.\ \cite{BMMS97} following the approach of Ref.\ \cite{SW74}. Locally, these
gauge transformations are the 4-parameter group of the inhomogeneous Lorentz
transformations. Up to second order in the perturbations an infinitesimal
change of coordinates 
\begin{equation}
\widetilde{x}^\mu \rightarrow x^\mu +\varepsilon \xi _{(1)}^\mu +\frac 12%
\varepsilon ^2(\xi _{(1);\nu }^\mu \xi _{(1)}^v+\xi _{(2)}^\mu ),
\label{CGT}
\end{equation}
where $\xi _{(1)}^\mu $ and $\xi _{(2)}^\mu $ are two independent arbitrary
vector fields and $\varepsilon $ a small (perturbative)
parameter, produces the following
effect on the first and second order perturbations of any quantity ${\bf %
\Phi }$ (scalar, vector or tensor field) that we assume can be expanded
as ${\bf \Phi^{(0)}}+{\bf \Phi^{(1)}}+{\bf \Phi^{(2)}}+...$
\begin{eqnarray}
&&\widetilde{{\bf \Phi }}^{(1)}\rightarrow {\bf \Phi }^{(1)}+{\cal L}_{\xi
_{(1)}}{\bf \Phi }^{(0)},  \label{CGI1} \\
&&\widetilde{{\bf \Phi }}^{(2)}\rightarrow {\bf \Phi }^{(2)}+{\cal L}_{\xi
_{(1)}}{\bf \Phi }^{(1)}+\frac 12 ({\cal L}_{\xi _{(1)}}^2
+{\cal L}_{\xi _{(2)}}){\bf \Phi }^{(0)}  \label{CGI2}
\end{eqnarray}
where, for the sake of completeness we recall here explicitly the basic
coordinate expressions of the Lie derivative along a vector field $\xi^\mu,$ 
\begin{eqnarray}
&&{\cal L}_{\xi}\Phi =\Phi _{,\mu }\xi ^\mu \text{ , if }\Phi \text{
is a scalar;}  \nonumber \\
&&{\cal L}_{\xi}\Phi ^\nu =\Phi _{,\mu }^\nu \xi ^\mu -\xi _{,\mu
}^\nu \Phi ^\mu \text{ , if }\Phi ^\nu \text{ is a vector;}  \label{LIE} \\
&&{\cal L}_{\xi}\Phi _{\alpha \beta }=\Phi _{\alpha \beta ,\mu }\xi
^\mu +\xi _{,\alpha }^\mu \Phi _{\mu \beta }\text{ }+\xi _{,\beta }^\mu \Phi
_{\alpha \mu }\text{ , if }\Phi _{\alpha \beta }\text{ is a tensor.} 
\nonumber
\end{eqnarray}

Note that from transformation (\ref{CGI1}) it follows that all 
Newman-Penrose quantities that vanishes on the background 
(or more precisely satisfy
${\cal L}_{\xi _{(1)}}{\bf \Phi }^{(0)}=0)$ , like 
$\psi _0^{(1)},$ $\psi _4^{(1)},
\psi_3^{(1)},k^{(1)},\sigma ^{(1)},\lambda ^{(1)},\nu ^{(1)}$, are 
first order gauge invariant (GI). 
Transformation (\ref{CGI2}), however, states that none
of these Newman-Penrose quantities, to the second order in the 
perturbations, are gauge invariant, since 
${\cal L}_{\xi _{(1)}}{\bf \Phi }^{(1)}\neq 0$. Thus, none of the
interesting Newman-Penrose quantities that are tetrad invariant (TI) 
and gauge invariant to the first order are also invariant to the second 
order in the perturbations. In particular, second order gauge invariance
requires that the quantity vanishes to zeroth and to first perturbative order.

Explicitly, for the scalar field $\psi_4$ we have 
\begin{equation}
\widetilde{\psi _4^{(2)}}\rightarrow \psi _4^{(2)}+\frac{\partial \psi
_4^{(1)}}{\partial x^\mu }\xi _{(1)}^\mu .  \label{CGT2}
\end{equation}
Hence we see that the vanishing of $\psi _4^{(0)}$ ensures that $\psi
_4^{(2)}$ will be gauge invariant under ``pure'' second order changes 
of coordinates, but since $\psi _4^{(1)}$ will in general depend on all 
four coordinates, $\psi_4^{(2)}$ will {\it not} be gauge invariant 
under {\it first} order changes of the coordinates.

In order to apply similar techniques to those we used to construct a
tetrad invariant object now in the coordinates context, i. e. by ``correcting''
$\psi_4^{(2)}$ with products of first order quantities, we will make use of
the following Lemma

\bigskip

{\bf Lemma:}\ {\it The product of the first order pieces $T^{(1)}P^{(1)}$
of two tensors (that can be expanded into perturbations) transforms
under a first plus second order gauge change, given by Eq.\ (\ref{CGT}),
as the product of the first order transformed quantities individually.}
$$
\widetilde{(T^{(1)}P^{(1)})} \rightarrow \left(T^{(1)}+{\cal L}_{\xi_{(1)}}
T^{(0)}\right)\left(P^{(1)}+{\cal L}_{\xi_{(1)}}
P^{(0)}\right)
$$

{\it Proof:} Let T and P be two general tensor fields. Apply the first 
plus second order transformation (\ref{CGI2})
to the product and consider second order pieces, then 
$$
\widetilde{(TP)}^{(2)} \rightarrow (TP)^{(2)}+{\cal L}_{\xi_{(1)}}
(TP)^{(1)}+ \frac 12 ({\cal L}_{\xi_{(1)}}^2
+{\cal L}_{\xi_{(2)}})(TP)^{(0)}, 
$$
or more explicitly
\begin{eqnarray}
\widetilde{(T^{(2)}P^{(0)}+T^{(1)}P^{(1)}+T^{(0)}P^{(2)})}\rightarrow 
&&(T^{(2)}P^{(0)}+T^{(1)}P^{(1)}+T^{(0)}P^{(2)}) 
+{\cal L}_{\xi_{(1)}}(T^{(1)}P^{(0)}+T^{(0)}P^{(1)})\nonumber\\
&&+\frac 12 ({\cal L}_{\xi _{(1)} }^2
+{\cal L}_{\xi _{(2)}})(T^{(0)}P^{(0)}).\nonumber
\end{eqnarray}

We now apply the same transformation
(\ref{CGI2}) to the products $P^{(0)}T$ and $T^{(0)}P$ to obtain 
$$
\widetilde{P^{(0)}T^{(2)}}\rightarrow P^{(0)}T^{(2)}+P^{(0)}{\cal L}%
_{\xi_{(1)}}(T^{(1)})+\frac 12 P^{(0)}({\cal L}_{\xi _{(1)}}^2
+{\cal L}_{\xi _{(2)}})(T)^{(0)}
$$
similarly 
$$
\widetilde{T^{(0)}P^{(2)}}\rightarrow T^{(0)}P^{(2)}+T^{(0)}{\cal L}%
_{\xi_{(1)}}(P^{(1)})+\frac 12 T^{(0)}({\cal L}_{\xi _{(1)}}^2
+{\cal L}_{\xi _{(2)}})(P)^{(0)}.
$$

Upon subtraction of the last two expressions from the first one, we obtain 
\begin{equation}
\widetilde{T^{(1)}P^{(1)}}\rightarrow T^{(1)}P^{(1)}+T^{(1)}{\cal L}_{\xi
_{(1)}}(P^{(0)})+P^{(1)}{\cal L}_{\xi _{(1)}}(T^{(0)})+
{\cal L}_{\xi _{(1)}}(T^{(0)}){\cal L}_{\xi _{(1)}}(P^{(0)}).
\label{uno+uno}
\end{equation}
This proves our Lemma. An obvious corollary is the case when both fields
are gauge invariant, i.e. ${\cal L}_{\xi _{(1)}}(T^{(0)})=0$ and 
${\cal L}_{\xi_{(1)}}(P^{(0)})=0$ this generates a second order 
quantity that is first and second order gauge (coordinate) invariant. 
\bigskip

To construct a second order gauge invariant waveform 
$\psi_{4\ GI}^{(2)}$ we can then use the same techniques as in the
previous subsection.
It is convenient now to start from our tetrad invariant object, as defined
in Eq.\ (\ref{invariante}). Under a first order coordinates change
$\psi_{4\ TI}^{(2)}$ transform as
\begin{equation}
\widetilde{\psi_{4\ TI}^{(2)}}\rightarrow \psi_{4\ TI}^{(2)}+
\psi_{4~,\mu}^{(1)}\xi_{(1)}^\mu+2\psi_4^{(1)} \bigg(\frac{l_{,\mu
}^{r~(0)}\xi_{(1)}^\mu-\xi_{(1),\mu}^r l^{\mu~(0)} }{l^{r~(0)}} +
\frac{m_{,\mu}^{\vartheta~(0)}\xi_{(1)}^\mu -\xi_{(1),\mu}^\vartheta
m^{\mu~(0)}}{m^{\vartheta~(0)}} \bigg)\label{VTI}
\end{equation}
where we made use of the properties expressed in Eqs \ (\ref{CGT2})
and \ (\ref{uno+uno}). 

As in Section III.A, the idea here is to add to $\psi_{4\ TI}^{(2)}$
terms quadratic in the first order perturbations in order to make
the whole object coordinate invariant\footnote{A similar procedure
was adopted to generate second order gauge invariants in the
Moncrief's formulation of Schwarzschild black hole perturbations\cite{GP98}.}
while preserving its tetrad invariance.
The procedure can be summarized as follows,

{\bf Prescription:} The first step is to invert the coordinate
transformations of first order quantities for the gauge vectors
$\xi_{(1)}^\mu$. We shall denote this first order
combination by the {\it boldface} vector:
${\mbox{\boldmath $\xi$}}_{(1)}^\mu$,
i.e. $\xi_{(1)}^\mu=\widetilde{\mbox{\boldmath $\xi$}_{(1)}^\mu}-
{\mbox{\boldmath $\xi$}}_{(1)}^\mu$.
Making the replacement $\xi_{(1)}^\mu\rightarrow
-\mbox{\boldmath $\xi$}_{(1)}^\mu$ into Eq.\ (\ref{VTI}) above generates
a totally invariant object. Still from all the possible invariant
objects we want those whose quadratic term do not contribute to the
radiation in an asymptotically flat gauge (AF). As we discussed at the end
of Sec.\ IV.A, this ensures us a simple interpretation of the 
invariant $\psi_I$ regarding radiated energy and waveforms.
Since Eq.\ (\ref{VTI}) is linear in $\xi_{(1)}^\mu$, subtracting
the quadratic term in an asymptotically flat gauge will be equivalent
to make the following replacement $\xi_{(1)}^\mu\rightarrow
\mbox{\boldmath $\xi$}_{(1)}^{\mu\ AF}-
{\mbox{\boldmath $\xi$}}_{(1)}^\mu$.
As we discussed before, a practical way to evaluate
$\mbox{\boldmath $\xi$}_{(1)}^{AF\mu}$ and keep the tetrad and coordinate
invariance is to use the outgoing radiation gauge (Eq.\ (\ref{outmetrica}))
and consider the final expression in terms of $\Psi_{ORG}$ as a
totally invariant expression regardless its derivation with
a choice of the first order gauge and tetrad.

We recall here
that $\psi_4^{(2)}$ and of course also terms quadratic in the first
order perturbations are already invariant under pure second order coordinate
transformations, labeled by $\xi_{(2)}^\mu$. Finally, our invariant waveform
can then be symbolically expressed as
\begin{eqnarray}
\psi_{4\ I}^{(2)}\dot=\psi_{4\ TI}^{(2)}+\psi_{4~,\mu}^{(1)}
\left({\mbox{\boldmath ${\mbox{\boldmath $\xi$}}$}}_{(1)}^{\mu\ ORG}-
{\mbox{\boldmath ${\mbox{\boldmath $\xi$}}$}}_{(1)}^\mu\right)
&&+2\psi_4^{(1)}\Biggr[\frac{l_{,\mu }^{r\ (0)}
\left({\mbox{\boldmath ${\mbox{\boldmath $\xi$}}$}}_{(1)}^{\mu\ ORG}-
{\mbox{\boldmath ${\mbox{\boldmath $\xi$}}$}}_{(1)}^\mu\right)
-l^{\mu\ (0)}
\left({\mbox{\boldmath ${\mbox{\boldmath $\xi$}}$}}_{(1),\mu}^{r\ ORG}-
{\mbox{\boldmath ${\mbox{\boldmath $\xi$}}$}}_{(1),\mu}^r\right)}
{l^{r~(0)}}\nonumber\\
&&+\frac{m_{,\mu}^{\vartheta~(0)}
\left({\mbox{\boldmath ${\mbox{\boldmath $\xi$}}$}}_{(1)}^{\mu\ ORG}-
{\mbox{\boldmath ${\mbox{\boldmath $\xi$}}$}}_{(1)}^\mu\right)
-m^{\mu~(0)}\left({\mbox{\boldmath ${\mbox{\boldmath $\xi$}}$}}_{(1),\mu}^
{\vartheta\ ORG}-
{\mbox{\boldmath ${\mbox{\boldmath $\xi$}}$}}_{(1),\mu}^\vartheta\right)
} {m^{\vartheta~(0)}}
\Biggr]. \label{GIW}
\end{eqnarray}

\subsection{Construction of the second order invariant waveform}

The above prescription is conceptually very simple. However, in practice,
to find ${\mbox{\boldmath $\xi$}}^t_{~(1)}$ and
${\mbox{\boldmath $\xi$}}^\varphi_{~(1)}$
brings some technical
complications. The first remark is that the procedure is not unique. 
We have a big choice of first order objects (all Newman-Penrose 
quantities, metric, extrinsic curvature, etc) to build up 
${\mbox{\boldmath $\xi$}}_{(1)}^\mu$. In fact, one can
easily see that the ambiguity to generate an invariant waveform 
has to be present since one can always add products of first order 
invariant objects to generate a new second order invariant object. 
The requirement that the quadratic correction must not influence the
asymptotic behavior greatly reduces this ambiguity. In fact,
physical quantities such as the radiated
energy and observed waveform, defined in an asymptotically flat region,
are uniquely defined by this method, since the differences 
introduced by different asymptotically flat coordinates vanish with
a higher power of $r$.
We thus, give an explicit object in order to
be able to make comparisons with, for instance, full numerical 
results that directly compute the covariant object $\psi_4$.
Below we give a simple choice of ${\mbox{\boldmath $\xi$}}_{(1)}^\mu$,
in order to construct $\psi_{4\ I}^{(2)}$ that is valid for 
perturbations of Kerr black holes, i.e. $a\not=0$. 
In Appendix C we give another choice for the case of a Schwarzschild 
background.

In the rest of this subsection, to simplify the notation,
we drop the subscript $(1)$ from the first order gauge
vectors ${\mbox{\boldmath $\xi$}}^\mu$
since we will never refer to the second order gauge vectors.
The ${\mbox{\boldmath $\xi$}}^r$ and ${\mbox{\boldmath $\xi$}}^\vartheta$
components can be easily found from the variations of the
tetrad invariant Weyl scalar $\psi_2^{(1)}$, 
\begin{eqnarray}
\widetilde{\psi _2^{(1)}}\rightarrow \psi _2^{(1)}+
\xi^r\partial_r{\psi _2^{(0)}}+\xi^\vartheta\partial_\vartheta
{\psi _2^{(0)}},
\end{eqnarray}
and of its complex conjugate $\overline{\psi}_2^{(1)}$, 
\begin{equation}
{\mbox{\boldmath $\xi$}}^r= -\frac{1}{6M}\left[\frac{\overline{\psi}_2^{(1)}}{%
\overline{\rho}^4}+ \frac{\psi_2^{(1)}}{\rho^4}\right],  \label{XIr}
\end{equation}
\begin{equation}
{\mbox{\boldmath $\xi$}}^\vartheta= -\frac{1}{6M(ia\sin{\vartheta})} \left[%
\frac{\overline{\psi}_2^{(1)}}{\overline{\rho}^4}-
\frac{\psi_2^{(1)}}{\rho^4}\right].
\label{XIth}
\end{equation}

The same techniques cannot be straightforwardly applied to find the
other two components ${\mbox{\boldmath $\xi$}}^t$ and
${\mbox{\boldmath $\xi$}}^\varphi$.
The origin of the problem can be traced back to the fact that the Kerr
metric has two two killing vectors along $\partial_t$ and $\partial_\varphi$,
and thus one can never find local, first order quantities that vary with
$\xi^t$ or $\xi^\varphi$, but only with the derivatives of them.
Explicitly, using
the variations of the metric and extrinsic curvature components (which are
tetrad invariant quantities), we find (here background fields are unlabeled) 
\begin{eqnarray}
&&{\mbox{\boldmath $\xi$}}^t_{,t}= \frac{ g_{\varphi\varphi}(h_{tt}^{(1)}
+g_{tt,r}
{\mbox{\boldmath $\xi$}}^r +g_{tt,\vartheta}{\mbox{\boldmath $\xi$}}%
^\vartheta) -2g_{t\varphi}(h_{t\varphi}^{(1)}+g_{t\varphi ,r}
{\mbox{\boldmath $\xi$}}^r +g_{t\varphi ,\vartheta}
{\mbox{\boldmath $\xi$}}^\vartheta
+g_{t\varphi}{\mbox{\boldmath $\xi$}}^\varphi_{,\varphi}+ g_{tt}
{\mbox{\boldmath $\xi$}}^t_{,\varphi})} {2(g_{t\varphi}^2-g_{tt}g_{\varphi%
\varphi})},  \nonumber \\
&&{\mbox{\boldmath $\xi$}}^t_{,r}=\frac{ g_{\varphi\varphi}(h_{tr}^{(1)}
+g_{rr}{\mbox{\boldmath $\xi$}}^r_{,t}) -g_{t\varphi}(h_{r\varphi}^{(1)}
+g_{rr}{\mbox{\boldmath $\xi$}}^r_{,\varphi})} 
{g_{t\varphi}^2-g_{tt}g_{\varphi\varphi}},  \nonumber \\
&&{\mbox{\boldmath $\xi$}}^t_{,\vartheta}= -\frac{(h_{t\vartheta}^{(1)}+g_{%
\vartheta\vartheta} {\mbox{\boldmath $\xi$}}^\vartheta_{,t}+g_{t\varphi}
{\mbox{\boldmath $\xi$}}^\varphi_{,\vartheta})}{g_{tt}} , \nonumber \\
&&{\mbox{\boldmath $\xi$}}^t_{,\varphi}= -\frac{(h_{\varphi\varphi}^{(1)}+g_{%
\varphi\varphi ,r} {\mbox{\boldmath $\xi$}}^r+g_{\varphi\varphi ,\vartheta}
{\mbox{\boldmath $\xi$}}^\vartheta +2g_{\varphi\varphi}
{\mbox{\boldmath $\xi$}}^\varphi_{,\varphi})} {2g_{t\varphi}},  \label{Xt}
\end{eqnarray}
and
\begin{eqnarray}
&&{\mbox{\boldmath $\xi$}}^\varphi_{,t}= \frac{2g_{tt}(h_{t\varphi}^{(1)}+g_{t%
\varphi ,r}{\mbox{\boldmath $\xi$}}^r +g_{t\varphi ,\vartheta}
{\mbox{\boldmath $\xi$}}^\vartheta +g_{t\varphi}{\mbox{\boldmath $\xi$}}%
^\varphi_{,\varphi}+ g_{tt}{\mbox{\boldmath $\xi$}}^t_{,\varphi})
-g_{t\varphi}(h_{tt}^{(1)}+g_{tt,r}{\mbox{\boldmath $\xi$}}^r +g_{tt,\vartheta}
{\mbox{\boldmath $\xi$}}^\vartheta)}
{2(g_{t\varphi}^2-g_{tt}g_{\varphi%
\varphi})},  \nonumber \\
&&{\mbox{\boldmath $\xi$}}^\varphi_{,r}= \frac{g_{tt}(h_{r\varphi}^{(1)}+
g_{rr}{\mbox{\boldmath $\xi$}}^r_{,\varphi})
-g_{t\varphi}(h_{tr}^{(1)}+g_{rr}{\mbox{\boldmath $\xi$}}^r_{,t})}
{g_{t\varphi}^2-g_{tt}g_{\varphi\varphi}} ,
\nonumber \\
&&{\mbox{\boldmath $\xi$}}^\varphi_{,\vartheta}= -\frac{K^{(1)}_{\vartheta%
\vartheta}}{2K_{\varphi\vartheta}},  \nonumber \\
&&{\mbox{\boldmath $\xi$}}^\varphi_{,\varphi}= -\frac{(K^{(1)}_{r\varphi}
+K_{\varphi r ,r}{\mbox{\boldmath $\xi$}}^r +
K_{\varphi r ,\vartheta}{\mbox{\boldmath $\xi$}}^\vartheta
+K_{\varphi r}{\mbox{\boldmath $\xi$}}^r_{,r})} {K_{\varphi r}}.
\label{Xphi}
\end{eqnarray}

Thus, to find ${\mbox{\boldmath $\xi$}}^t$ and
${\mbox{\boldmath $\xi$}}^\varphi$ one has to integrate their four
derivatives over the spacetime. This can be performed like the integration
of a potential in four dimensions. For that one has to verify the integrability
conditions. In practice, since we are going to compute differences of these
vectors, with respect to the asymptotically flat ones, the existence of
the $\xi^t$ and $\xi^\varphi$ components are assumed a priori and they
are related to the existence of the outgoing radiation gauge proved
in Ref.\ \cite{C75}.
As a consequence of these integrals on first order fields,
the resulting waveform
will be nonlocal, but this carries no further consequences since
in solving the second order perturbations
we assumed first order ones to be completely known. Notably, the
evolution equation for the second order perturbations is {\it local}. In
fact, only derivatives of ${\mbox{\boldmath $\xi$}}^t$ and
${\mbox{\boldmath $\xi$}}^\varphi$ enter in building
up the source
\begin{equation}
\widehat{{\cal T}}[\psi_{I}^{(2)}]=S_{I},\quad
\psi_{I}^{(2)}\dot = (\rho^{(0)})^{-4}\psi_{4\ I}^{(2)},
  \label{final}
\end{equation}
where the source term (as can be derived from Eq. (\ref{unodos})) is now
\begin{eqnarray}
S_{I}&=& 2(\rho^{(0)})^{-4}
\Sigma \left\{{\cal S}_4[\psi_4^{(1)}]+T[\text{matter}%
]\right\} \\
&&+\widehat{{\cal T}}\left[2\psi^{(1)}\left(\frac{l^{r~(1)}-l^{r~(1)}_{ORG}}
{l^{r~(0)}}+ \frac{m^{\vartheta~(1)}-m^{\vartheta~(1)}_{ORG}}
{m^{\vartheta~(0)}}\right) -\frac23 
\frac{\left(\psi_3^{(1)}-\psi_3^{(1)\ ORG}\right)^2}
{(\rho^{(0)})^{4}\psi_2^{(0)}}\right]
\nonumber \\
&+&\widehat{{\cal T}}\left[ \psi_{,\mu}^{(1)}
\xi^\mu_{~(1)}+
2\psi^{(1)} \bigg(\frac{l_{,\mu }^{r\ (0)}\xi^\mu_{~(1)} 
-\xi_{,\mu~(1)}^r l^{\mu\ (0)}}{l^{r~(0)}} +\frac{
m_{,\mu}^{\vartheta~(0)}\xi^\mu_{~(1)} -
\xi_{,\mu~(1)}^\vartheta m^{\mu~(0)}} {m^{\vartheta~(0)}}
\bigg)\right],  \label{GIsource}
\end{eqnarray}
here $\xi^\mu\dot=
{\mbox{\boldmath ${\mbox{\boldmath $\xi$}}$}}_{(1)}^{\mu\ ORG}-
{\mbox{\boldmath ${\mbox{\boldmath $\xi$}}$}}_{(1)}^\mu$.

While the evolution is local, we need to compute the waveform at least on
the initial hypersurface and then at the observer location (to compute,
for instance, the radiated energy). At $t=0$, after mode decomposition
in the $\varphi$ coordinate, we have
\begin{equation}
\xi^i=\sum_{\text{m}\not=0}\frac{({\mbox{\boldmath$\xi$}}^i_{,\varphi}
-{\mbox{\boldmath$\xi$}}^{i\ ORG}_{,\varphi}
)_{\text{m}}e^{i\text{m}\varphi}}{i\text{m}}+
\int dr({\mbox{\boldmath $\xi$}}^i_{,r}
-{\mbox{\boldmath $\xi$}}^{i\ ORG}_{,r})
+\int d\vartheta({\mbox{\boldmath $\xi$}}^i_{,\vartheta}
-{\mbox{\boldmath $\xi$}}^{i\ ORG}_{,\vartheta})
-\int drd\vartheta({\mbox{\boldmath $\xi$}}^i_{,r\vartheta}
-{\mbox{\boldmath $\xi$}}^{i\ ORG}_{,r\vartheta})+c^i,
\end{equation}
where $i=(t,\varphi)$ and the same equation holds for the observer at
a fixed $r_{obs}$, exchanging the roles of $r$ and $t$.

Note the presence of the integration constants $c^t$ and $c^\varphi$.
They represent first order changes in the origin of time and azimuthal angle.
This problem was already found in Ref.\ \cite{GNPP98} and there it was
given a method to fix ``a posteriori'' the value of the constants.
In Section IV.B we generalize the procedure given in Ref.\ \cite{GNPP98}
and explicity write the integrals that are necessary to fix the 
constants $c^t$ and $c^\varphi$.

In order to compute the totally invariant second order 
waveform $\psi_{4\ Ic}^{(2)}$ 
we must fix the constants $c_t$ and $c_\varphi$.
We can generalize the gauge fixing 
prescription given in Ref.\cite{NGPP98} and define
``a posteriori''' the value of the constants
\begin{equation}
c^t=\frac{\int_{-\infty}^{\infty}dt\dot\psi_4^{(1)}\psi_{4\ I}^{(2)}
\int_{-\infty}^{\infty}dt(\partial_\varphi\psi_4^{(1)})^2-
\int_{-\infty}^{\infty}dt\partial_\varphi\psi_4^{(1)}\psi_{4\ I}^{(2)}
\int_{-\infty}^{\infty}dt\dot\psi_4^{(1)}\partial_\varphi\psi_4^{(1)}
}
{
\int_{-\infty}^{\infty}dt(\dot\psi_4^{(1)})^2
\int_{-\infty}^{\infty}dt(\partial_\varphi\psi_4^{(1)})^2-
\int_{-\infty}^{\infty}dt\dot\psi_4^{(1)}\partial_\varphi\psi_4^{(1)}
\int_{-\infty}^{\infty}dt\dot\psi_4^{(1)}\partial_\varphi\psi_4^{(1)}
},  \label{ct}
\end{equation}
\begin{equation}
c^\varphi=\frac{
\int_{-\infty}^{\infty}dt\dot\psi_4^{(1)}\partial_\varphi\psi_4^{(1)}
\int_{-\infty}^{\infty}dt\dot\psi_4^{(1)}\psi_{4\ I}^{(2)}-
\int_{-\infty}^{\infty}dt\partial_\varphi\psi_4^{(1)}\psi_{4\ I}^{(2)}
\int_{-\infty}^{\infty}dt(\dot\psi_4^{(1)})^2
} 
{
\int_{-\infty}^{\infty}dt(\dot\psi_4^{(1)})^2
\int_{-\infty}^{\infty}dt(\partial_\varphi\psi_4^{(1)})^2-
\int_{-\infty}^{\infty}dt\dot\psi_4^{(1)}\partial_\varphi\psi_4^{(1)}
\int_{-\infty}^{\infty}dt\dot\psi_4^{(1)}\partial_\varphi\psi_4^{(1)}
},\label{cphi}
\end{equation}
We can then construct the ``c-invariant'' waveform
\begin{equation}
\psi_{Ic}^{(2)}=\psi_{I}^{(2)}-(c^t\partial_t+
c^\varphi\partial_\varphi)\psi^{(1)},  \label{PW}
\end{equation}
This procedure amounts to gauge fixing the zero of time and of the azimuthal
angle in such a way that
the integrals in the numerators of Eqs. \ (\ref{ct}) and \ (\ref{cphi})
vanish. In order to be able to compare the perturbative results with the
full numerically ones it is crucial that one is able to perform
the same origin of coordinates fixing.
Note that we can also fix these constants at the initial hypersurface $t=0$.
The same expressions (\ref{ct})-(\ref{cphi}) apply changing the integrations
in time by integrations in $r$.

\section{Summary and discussion}

In this paper we presented a gauge and tetrad invariant framework
for studying the evolution of general second order
perturbations about a rotating black hole.
To do so, we first uncoupled second (and higher) order perturbations of
Kerr black holes for the Weyl scalar $\psi_4$, that directly
represents the outgoing
gravitational radiation, and found that the perturbed outgoing radiation
field $\psi_4^{(n)}$ fulfils a single Teukolsky-like equation (see Eq.\ (\ref
{dos})) with the same wave operator as for the first order perturbations\cite
{T73}, acting on the left hand side and an additional source term written
as products of lower order perturbations on the right hand side
of the equation. We note, however, that $\psi_4^{(2)}$ is
neither tetrad nor {\it first order}
coordinate (gauge) invariant. It is only invariant
under purely {\it second}
order changes of coordinates, simply because $\psi_4$
vanishes on the background (Kerr metric). Invariant objects to describe
second perturbations lead us to reliable physical
answers without having to face gauge difficulties. Hence, we explicitly
show that it is always possible to correct $\psi_4^{(2)}$ in order to
build up a complete second order invariant waveform
$\psi_{I}^{(2)}$ (i. e. invariant under both tetrad rotations and
infinitesimal coordinates transformations) that gives a measure of the
outgoing gravitational radiation. This is done in Sec.\ IV where we
give a general {\it prescription} to produce the result
expressed in Eq.\ (\ref{GIW}). We also show that the same equation as
(\ref{dos}), with a ``corrected'' source term, is now satisfied by
$\psi_{I}^{(2)}$
(see Eq.\ (\ref{final})). A number of interesting conceptual and
technical issues raised from this computation, like the appearance of
nonlocalities in the definition of the gauge invariant waveform when
we want to relate it to known first order objects and 
its non uniqueness. Seen in retrospective, our method of generating
a gauge invariant object is like a machine that transforms any
(first order) gauge into an asymptotically flat one,
in particular, into the outgoing radiation gauge. In fact,
in this gauge we have $(\psi_4^{(2)})^{ORG}=(\psi_{4L}^{(2)})^{ORG}=
(\psi_I^{(2)})^{ORG}=\psi_I^{(2)}$. This fits into Bardeen's\cite{B80}
interpretation of a gauge independent quantity and suggest to work in
the outgoing radiation gauge as a particularly simple way of dealing
with the numerical integration of the second order equations\cite{CKL99}.
In the language of Eq.\ (\ref{CGI2}) we see that the process of building
up $\psi_I$ is like subtracting the first order piece to $\psi_4$.
Our gauge invariant object, $\psi_I$, is not the second order term of a
series expansion of $\psi_4$, but it can be related to $psi_4^{(2)}$
in an asymptotically flat gauge.

The spirit of this work has been to show that there exists a gauge invariant
way to deal with second order perturbations in the more general case of a
rotating black hole and to provide theoretical support to the numerical
integration of the second order perturbation problem. In order to
implement such integration of Eq.\ (\ref{final}) we proceed as
follows: We assume that on an initial hypersurface we know the first and
second order perturbed metric and extrinsic curvature. We then solve the
first order problem, i.e. solve the standard Teukolsky equation for
$\psi_4^{(1)}$ (and for $\psi_0^{(1)}$).
Next we build up the perturbed metric coefficients in, for
instance, the outgoing radiation gauge\ (\ref{outcalibre}). The perturbed
spin coefficients are now given by expression\ (\ref{A4}) and the perturbed
covariant basis by\ (\ref{A2}). Those are all the necessary elements to build
up the effective source term appearing on the right hand side of our
evolution equation, as explained in Section III.A.
It is worth to note here that from the analysis of the asymptotic behavior
of the different Newman-Penrose quantities\cite{NP62} involved in the 
source, one can see that at infinity the envelope of (the oscillating) 
$S_4$ is at least of ${\cal O}(r^{-2})$, which guarantees 
the convergence of the integration of Eq.\ (\ref{final}). 

The other piece of information that we need in order to integrate
Eq.\ (\ref{final}) is $\psi_{I}^{(2)}$ on the initial
hypersurface. This is explained in Section III.B. We also need to use in this
case Eq.\ (\ref{segundo}) and the expressions given in Appendix B.
For the computation of the radiated energy and momentum one uses
Eqs.\ (\ref{energy}) and (\ref{momentum}).
The advantage of this procedure is that we can now use the same
(2+1)-dimensional code
for evolving the first order perturbations\cite{KLPA97} by adding
a source term.
In fact, the background (Kerr) metric allows a decomposition into axial modes,
i.e. the variable $\varphi$. A mode decomposition of all quantities
involved in the second order evolution equation can be trivially performed
(note that in the source, involving quadratic terms in the first order
perturbations one has to include a double sum over modes, let us say,
$\text{m}$ and $\text{m}^{\prime}$). In the time domain no further mode 
decomposition (i.e. in $\ell-$multipole) of the source term is practical.

An important application of the formalism presented in this paper\cite{CKL99}
is to reproduce the results obtained in Refs.\ \cite{GNPP96}
(for the nonrotating case and the multipole $\ell=2$).
The complexity of the calculations in the standard Zerilli formalism that
would follow from considering the sum over all multipoles can be notably
simplified in the Newman-Penrose formalism. We can thus also
study the $\ell=4$ multipole of the radiation and
not only test the
efficiency of our formalism, but also make a more detailed
comparison with full numerical results.
The next step is to extend the numerical computation to the more
interesting case of rotating black holes. The numerical integration of
Eq.\ (\ref{final}) will be relevant not only for
establishing the range of validity of the collision parameters
in the close limit approximation, but (hopefully) to produce a more
precise computation of the gravitational radiation. Direct
comparison with the existing codes for numerical integration of the
full nonlinear Einstein equations is possible\cite{BS95}.

Following the steps described in this paper, upon exchange of the null
directions $l\leftrightarrow n$ and $\overline{m}\leftrightarrow m$, it is
straightforward to write the corresponding
equations for $\psi_0$ in case
one wants to have a description in terms of ingoing
waves. This would allow to study the influence of gravitational
radiation on the horizon of a rotating black hole,
critical collapse and also
phenomena in their interior, like the mass-inflation.
We studied in detail only gravitational perturbations, but it seems
straightforward to generalize our method to scalar and vector perturbations.
We also note that, although we have focused our
attention on the problem of colliding black holes, the second
order perturbative formalism developed in this paper can be easily
generalized to any Petrov type D (or even type II) background metric and
thus can be applied to study other interesting astrophysical
scenarios as nonrotating neutron stars and cosmology.

\begin{acknowledgments}
We thank M. Bruni, A. Garat, W. Krivan, R. H. Price, J. Pullin 
B. Schmidt and B. Whiting for many fruitful 
comments and reading the original manuscript.
The authors acknowledges Deutsche Forschungsgemeinschaft SFB 382 
for partial financial support.
C.O.L. is a member of the Carrera del Investigador Cient\'\i fico 
of CONICET, Argentina
and thanks FUNDACI\'ON ANTORCHAS for partial financial support.
\end{acknowledgments}

\appendix 

\section{First order Newman-Penrose quantities}

Throughout this Appendix, to simplify the notation, we omit the superscript
$(0)$ on the background quantities, while all first order quantities are
denoted with the superscript $(1)$ with the exception of the first order
metric perturbation that we simply denote as $h_{\mu \nu \text{.}}$

Let us first note that the perturbed null tetrad can be represented by \cite
{C76} 
\begin{eqnarray}
l_{\ }^{\mu (1)} &=&-\frac 12h_{ll}n^\mu ,  \nonumber \\
n^{\mu (1)} &=&-\frac 12h_{nn}l^\mu -h_{nl}n^\mu ,  \label{A1} \\
m^{\mu (1)} &=&\frac 12h_{mm}\overline{m}^\mu +\frac 12h_{m\overline{m}%
}m^\mu -h_{ml}n^\mu -h_{mn}l^\mu .  \nonumber
\end{eqnarray}
Note that in order to have this explicit form a choice of the first order
null directions was made. To relate this to the metric perturbation recall
that $g_{\mu \nu }=2l_{(\mu }n_{\nu )}-2m_{(\mu }\overline{m}_{\nu)}$ which
implies that $h_{\mu \nu }=
2l_{(\mu }^{(1)}n_{\nu )}+2l_{(\mu }n_{\nu)}^{(1)}
-2m_{(\mu }^{(1)}\overline{m}_{\nu )}-2m_{(\mu }\overline{m}_{\nu )}^{(1)}$.

Making use of the relations (\ref{A1}) we can immediately derive the first
order Newman-Penrose directional derivatives

\begin{eqnarray}
D^{(1)} &=&l_{\ }^{\mu (1)}\partial_\mu =-\frac 12h_{ll}\Delta ,  \nonumber
\\
\Delta^{(1)} &=&n_{\ }^{\mu (1)}\partial_\mu =-\frac 12h_{nn}D-h_{nl}\Delta ,
\label{A2} \\
\delta^{(1)} &=&m_{\ }^{\mu (1)}\partial_\mu =\frac 12h_{mm}\overline{\delta 
}+\frac 12h_{m\overline{m}}\delta -h_{ml}\Delta -h_{mn}D.  \nonumber
\end{eqnarray}

In order to compute the spin coefficients to the required order we follow
Ref.\ \cite{C76} making use of the commutation relations \cite{C83}, Ch. 1.8 
(these are exact expressions)
\begin{eqnarray}
\Delta D-D\Delta &=&\left( \gamma +\overline{\gamma }\right) D+\left(
\epsilon +\overline{\epsilon }\right) \Delta -\left( \overline{\tau }+\pi
\right) \delta -\left( \tau +\overline{\pi }\right) \overline{\delta }, 
\nonumber \\
\delta D-D\delta &=&\left( \overline{\alpha }+\beta -\overline{\pi }\right)
D+\kappa \Delta -\left( \overline{\rho }+{\epsilon }-\overline{\epsilon}
\right) \delta -\sigma \overline{\delta },  \label{A3} \\
\delta \Delta -\Delta \delta &=&-\overline{\nu }D+\left( \tau -\overline{%
\alpha }-\beta \right) \Delta +\left( \mu +\overline{\gamma }-\gamma \right)
\delta -\overline{\lambda }\overline{\delta },  \nonumber \\
\overline{\delta }\delta -\delta \overline{\delta } &=&\left( \overline{\mu }%
-\mu \right) D+\left( \overline{\rho }-\rho \right) \Delta +\left( \alpha -%
\overline{\beta }\right) \delta +\left( \beta -\overline{\alpha }\right) 
\overline{\delta },  \nonumber
\end{eqnarray}
expanding both sides to first perturbative order and using Eq.\ (\ref{A2})
we can equate the coefficients of each operator to get a system of linear
equations (16 of which only 12 are independent) that can be solved for the
spin coefficients giving

\begin{eqnarray}
k^{(1)} &=&\left( D-\overline{\rho }-2\ \epsilon \right) h_{lm}-\frac 12%
\left( \delta -2\alpha -2\beta +\overline{\pi }+\tau \right) h_{ll}, 
\nonumber \\
\sigma^{(1)} &=&\left( \overline{\pi }+\tau \right) h_{lm}+\frac 12\left(
D+\rho -\overline{\rho }+2\overline{\epsilon }-2\epsilon \right) h_{mm}, 
\nonumber \\
\nu^{(1)} &=&-\left( \Delta +\overline{\mu }+2\gamma \right) h_{n\overline{m}%
}+\frac12 \left( \overline{\delta }+2\alpha +2\overline{\beta }-\pi -%
\overline{\tau }\right) h_{nn},  \nonumber \\
\lambda^{(1)} &=&-\left( \overline{\tau }+\pi \right) h_{n\overline{m}}-%
\frac 12\left( \Delta +\overline{\mu }-\mu +2\gamma -2\overline{\gamma }%
\right) h_{\overline{m}\overline{m}},  \nonumber \\
2\mu^{(1)} &=&\rho h_{nn}-\left( \delta +2\beta +\tau \right) h_{n\overline{m%
}}+\left( \overline{\delta }+2\overline{\beta }-2\pi -\overline{\tau }%
\right) h_{nm}-\frac 12\left( 2\Delta +\overline{\mu }-\mu +\gamma -%
\overline{\gamma }\right) h_{m\overline{m}}  \nonumber \\
2\rho^{(1)} &=&\overline{\mu }h_{ll}+\left( \rho -\overline{\rho }\right)
h_{nl}+\left( D+\rho -\overline{\rho }\right) h_{m\overline{m}}-\left(
\delta -2\overline{\alpha }-\overline{\pi }\right) h_{l\overline{m}}+\left( 
\overline{\delta }+2\overline{\tau }-2\alpha +\pi \right) h_{lm},  \nonumber
\\
2\epsilon^{(1)} &=&\left( D+\rho -\overline{\rho }\right) h_{nl}+\frac 12%
\left( \overline{\delta }-2\alpha -\pi \right) h_{lm}-\frac 12\left( \delta
-2\overline{\alpha }+3\pi +4\tau \right) h_{l\overline{m}}  \nonumber \\
&&+\frac 12 \left( \rho -\overline{\rho }\right) h_{m\overline{m}}-\frac 12%
\left( \Delta +2\gamma \right) h_{ll},  \label{A4} \\
2\pi^{(1)} &=&-\left( D-\rho -2\epsilon \right) h_{n\overline{m}}-\left( 
\overline{\delta }+\overline{\tau }+\pi \right) h_{nl}-\left( \Delta +%
\overline{\mu }-2\overline{\gamma }\right) h_{l\overline{m}}-\overline{\tau }%
h_{m\overline{m}}-\tau h_{\overline{m}\overline{m}},  \nonumber \\
2\tau^{(1)} &=&\left( D-\overline{\rho }+2\overline{\epsilon }\right)
h_{nm}+\left( \delta -\overline{\pi }-\tau \right) h_{nl}+\left( \Delta +\mu
-2\gamma \right) h_{lm}-\overline{\pi }h_{m\overline{m}}-\pi h_{mm}, 
\nonumber \\
2\alpha^{(1)} &=&\frac 12\left( D-2\overline{\rho }-\rho -2 \epsilon \right)
h_{n\overline{m}} -\frac 12\left( \Delta +4\gamma-2\mu + 
\overline{\mu }-2\overline{\gamma} \right) h_{l\overline{m}}  \nonumber \\
&&-\frac 12\left(\overline{ \delta} +\pi+\overline{\tau} \right) h_{nl}+%
\frac 12\left( \overline{\delta} +2\alpha-\pi-\overline{\tau} \right) h_{m%
\overline{m}}-\frac 12\left( \delta-2\overline{\alpha} +\overline{\pi}
+\tau\right) h_{\overline{m}\overline{m}},  \nonumber \\
2\beta^{(1)} &=&\frac 12\left( D-\overline{\rho }-4\epsilon +2\rho +2%
\overline{\epsilon }\right) h_{nm}-\frac 12\left( \Delta +\mu +2\overline{%
\mu }+2\gamma \right) h_{lm}  \nonumber \\
&&-\frac 12\left( \delta +\overline{\pi }+\tau \right) h_{nl}-\frac 12
\left( \delta -2\beta +\overline{\pi }+\tau \right) h_{m\overline{m}}+\frac 1%
2\left( \overline{\delta }+2\beta -\pi -\overline{\tau }\right) h_{mm}, 
\nonumber \\
2\gamma^{(1)} &=&-\left(\overline{\gamma }+\gamma \right) h_{nl}+\frac 12%
\left( D+\rho -\overline{\rho }+2\overline{\epsilon }\right) h_{nn}-\frac 12%
\left( \delta +2\beta +2\overline{\pi }+3\tau \right) h_{n\overline{m}} 
\nonumber \\
&&+\frac 12\left( \overline{\delta }+2\overline{\beta }-2\pi -\overline{\tau 
}\right) h_{nm}+\frac 14\left( 3\overline{\mu }-2\mu +\gamma -\overline{%
\gamma }\right) h_{m\overline{m}}  \nonumber
\end{eqnarray}

Note that these expressions are completely independent of the choice of the
gauge, although a tetrad choice to first order had to be made in Eq.\ (\ref
{A1}).

Finally, the {\it exact} Weyl scalars are

\begin{equation}
\psi_0=(D-3\epsilon +\overline{\epsilon }-\rho -\overline{\rho })\sigma
-(\delta -\overline{\alpha }-3\beta +\overline{\pi }-\tau )\kappa ,\nonumber
\end{equation}

\begin{eqnarray}
\psi _1 &=&(D+\overline{\epsilon }-\overline{\rho })\beta -(\delta -%
\overline{\alpha }+\overline{\pi })\epsilon -(\alpha +\pi )\sigma +(\gamma
+\mu )\kappa,\nonumber
\end{eqnarray}

\begin{eqnarray}
\psi _2 &=&\left[(\overline{\delta }-2\alpha +\overline{\beta }-\pi -%
\overline{\tau })\beta -(\delta -\overline{\alpha }+\overline{\pi } +\tau
)\alpha\right.  \nonumber \\
&&+(D+\epsilon +\overline{\epsilon }+\rho -\overline{\rho })\gamma -(\Delta
- \overline{\gamma }-\gamma +\overline{\mu }-\mu )\epsilon \nonumber\\
&&\left. +(\overline{\delta }-\alpha +\overline{\beta }-\overline{\tau }%
-\pi)\tau -(\Delta -\overline{\gamma }-\gamma +\overline{\mu }-\mu)\rho
+2(\nu \kappa -\lambda \sigma )\right]/3,  \nonumber
\end{eqnarray}

\begin{eqnarray}
\psi _3 &=&(\overline{\delta }+\overline{\beta }-\overline{\tau })\gamma
-\left( \Delta -\overline{\gamma }+\overline{\mu} \right) \alpha + (\epsilon
+\rho )\nu-(\beta +\tau )\lambda,\nonumber
\end{eqnarray}

and

\begin{equation}
\psi_4=(\overline{\delta }+3\alpha +\overline{\beta }+\pi -\overline{\tau }%
)\nu -(\Delta -\overline{\gamma }+3\gamma +\mu +\overline{\mu })\lambda ,
\label{psi4exacto}
\end{equation}
Note that these expressions can be trivially expanded to first perturbative
order and hold when matter sources are included.

\section{Second order Newman-Penrose quantities}

Taking the same first order choice of the tetrad to second order 
(we can do this because the final aim is to plug this into an invariant
object) one obtains 
\begin{eqnarray}
l_{\ }^{\mu (2)} &=&-\left[\frac 12h_{ll}^{(2)}-h_{ll}^{(1)}
h_{ln}^{(1)}+2h_{l\overline{ m}}^{(1)} h_{lm}^{(1)}\right]n^\mu ,  \nonumber
\\
n^{\mu (2)} &=&-\left[\frac 12h_{nn}^{(2)}+2h_{n\overline {m}%
}^{(1)}h_{nm}^{(1)}\right]l^\mu -\left[h_{nl}^{(2)}
-h_{ll}^{(1)}h_{nn}^{(1)}+\frac 12\left(h_{lm}^{(1)}h_{n\overline{ m}%
}^{(1)}+h_{l\overline{ m}}^{(1)}h_{nm}^{(1)}\right)\right]n^\mu,  \label{B1}
\\
m^{\mu (2)} &=&-\left[h_{mn}^{(2)}+h_{n\overline{m}}^{(1)} h_{mm}^{(1)}+h_{m%
\overline{m}}^{(1)}h_{nm}^{(1)}\right]l^\mu -\left[h_{ml}^{(2)}+h_{l%
\overline{m}}^{(1)} h_{mm}^{(1)}+h_{m\overline{m}}^{(1)}h_{lm}^{(1)}%
\right]n^\mu  \nonumber \\
&+&\frac 12\left[h_{m\overline{m}}^{(2)} +\frac12
h_{\overline{m}\overline{m}}^{(1)}
h_{mm}^{(1)}+\frac 12h_{m\overline{m}}^{(1)} h_{m\overline{m}%
}^{(1)}\right]m^\mu +\frac 12\left[h_{\overline{m}\overline{m}}^{(2)} +h_{m%
\overline{m}}^{(1)}h_{\overline{m}\overline{m}}^{(1)} \right]\overline{m}%
^\mu.  \nonumber
\end{eqnarray}

We now expand up to second order the third commutator of Eq.\ (\ref{A2}) to
obtain $\nu^{(2)}\dot=\nu_L^{(2)}+\nu_Q^{(2)}$ and $\lambda^{(2)}\dot=%
\lambda_L^{(2)}+\lambda_Q^{(2)}$ 
\begin{eqnarray}
\nu_L^{(2)}&=&-(\Delta+\overline{\mu}+2\gamma)h_{n\overline{m}}^{(2)} +\frac 
12(\overline\delta+\pi-\overline{\tau})h_{nn}^{(2)},  \nonumber \\
\nu_Q^{(2)}&=&-(\Delta+\overline{\mu}+2\gamma) h_{nm}^{(1)}h_{\overline{m}%
\overline{m}}^{(1)} +2(\overline\delta+\pi-\overline{\tau}) h_{nm}^{(1)}h_{n%
\overline{m}}^{(1)}-\lambda^{(1)}h_{nm}^{(1)}  \nonumber \\
&-&\frac 12\left[(\overline{\delta} +\overline{\tau}-\alpha-\beta)^{(1)}+(%
\gamma+\overline\gamma) h_{l\overline{m}}^{(1)}\right]h_{nn}^{(1)}-\left[
(\mu-\gamma+\overline\gamma)^{(1)}-(\gamma+\overline\gamma)
h_{lm}^{(1)}\right]h_{n\overline{m}}^{(1)},  \nonumber \\
\lambda_L^{(2)}&=&-\frac 12(\Delta+\overline{\mu}-\mu- 2\overline{\gamma}%
+2\gamma)h_{\overline{m}\overline{m}}^{(2)} -(\overline{\tau}+\pi)h_{n%
\overline{m}}^{(2)}, \\
\lambda_Q^{(2)}&=&-\frac 12(\Delta+\overline{\mu}-\mu- 2\overline{\gamma}%
+2\gamma)h_{\overline{m}\overline{m}}^{(1)} h_{\overline{m}m}^{(1)}+(%
\overline{\tau}+\pi) \left[h_{\overline{m}n}^{(1)}h_{ln}^{(1)} -h_{\overline{%
m}\overline{m}}^{(1)}h_{nm}^{(1)} -\frac 12 h_{\overline{m}%
l}^{(1)}h_{nn}^{(1)}\right]  \nonumber \\
&-&\frac 12 \lambda^{(1)}h_{\overline{m}m}^{(1)} -\frac12 
\left[(\overline{\mu}-%
\overline{\gamma}+\gamma)^{(1)} -\frac 12 \overline{\rho}h_{nn}^{(1)}+(\mu-%
\gamma+\overline\gamma) h_{ln}^{(1)}\right]h_{mm}^{(1)}.  \nonumber
\end{eqnarray}

Finally, using Eq.\ (\ref{psi4exacto}) we find $\psi _4^{(2)}\doteq \psi _{4%
\text{L}}^{(2)}+\psi _{4\text{Q}}^{(2)},$ 
\begin{eqnarray}
\psi _{4\text{L}}^{(2)}&=&(\overline{\delta}+3\alpha+\overline\beta+\pi-
\overline{\tau})\nu_L^{(2)}-(\Delta+\overline{\mu}+\mu- \overline{\gamma}%
+3\gamma)\lambda_L^{(2)},  \nonumber \\
\psi _{4\text{Q}}^{(2)}&=&(\overline{\delta}+3\alpha+\overline\beta+\pi-
\overline{\tau})\nu_Q^{(2)}-(\Delta+\overline{\mu}+\mu- \overline{\gamma}%
+3\gamma)\lambda_Q^{(2)} \\
&+&(\overline{\delta}+3\alpha+\overline\beta+\pi -\overline{\tau}%
)^{(1)}\nu^{(1)}-(\Delta+\overline{\mu}+\mu- \overline{\gamma}%
+3\gamma)^{(1)}\lambda^{(1)}.  \nonumber
\end{eqnarray}

\section{Gauge invariants in the Schwarzschild limit}

In the case when $a=0$, we can find the following set of 
{\it first} order gauge vectors
assuming that the ${\mbox{\boldmath $\xi$}}_{\text{m}}^r$ is given by
Eq.\ (\ref{XIr})

\begin{eqnarray}
{\mbox{\boldmath $\xi$}}_{\text{m}}^\vartheta&=& -
\frac{\cos\vartheta(2h_{\varphi\varphi}+2\sin^2\vartheta
r{{\mbox{\boldmath $\xi$}}^r_{\text{m}}} -\sin^2\vartheta
h_{\vartheta\vartheta})+\sin\vartheta (2i\text{m}
h_{\vartheta\varphi}-h_{\varphi\varphi,\vartheta} -2\sin^2\vartheta
r{\mbox{\boldmath $\xi$}}^r_{\text{m},\vartheta})} {2r^2\sin\vartheta(
\text{m}^{2}+1)}, \nonumber\\
i\text{m}{{\mbox{\boldmath $\xi$}}^t_{\text{m}}}&=& \frac{r}{2(r-3M)}
\{r h_{t\varphi,r}-2h_{t\varphi}+r^3\sin^2\vartheta 
{\mbox{\boldmath $\xi$}}^\varphi_{\text{m},rt}+i\text{m}(r-2M)
{\mbox{\boldmath $\xi$}}^t_{\text{m},r}\} \nonumber\\
{i\text{m}{\mbox{\boldmath $\xi$}}^\varphi_{\text{m}}}&=& -
\frac{1}{2r^2\sin^2\vartheta (\text{m}^2+1)} \{
(\text{m}^2+1-2\cos^2\vartheta)h_{\varphi\varphi} +
\cos^2\vartheta\sin^2\vartheta h_{\vartheta\vartheta} \nonumber\\
&+&\cos\vartheta\sin\vartheta (h_{\varphi\varphi ,\vartheta}
-2i\text{m}h_{\vartheta\varphi}+\sin^2\vartheta r
{\mbox{\boldmath $\xi$}}^r_{\text{m},\vartheta})
+2\sin^2\vartheta(\text{m}^2+\sin^2\vartheta)
{\mbox{\boldmath $\xi$}}^r_{\text{m}}\}, \nonumber\\
{\mbox{\boldmath $\xi$}}^t_{\text{m},\vartheta}&=& -
\frac{r}{\sin2\vartheta (r-2M)(r-3M)} \{-i\text{m}
r^2h_{t\varphi,r}+\text{m}^2r^2{\mbox{\boldmath $\xi$}}^t_{\text{m},r}
-4\text{m}^2Mr{\mbox{\boldmath $\xi$}}^t_{\text{m},r}
-i\text{m}r^4\sin^2\vartheta{\mbox{\boldmath $\xi$}}^{\varphi}_{\text{m},tr}
\nonumber\\
&+&2i\text{m}Mr h_{t\varphi,r}+4\text{m}^2M^2
{\mbox{\boldmath $\xi$}}^t_{\text{m},r}+2i\text{m}MR^3\sin^2\vartheta
{\mbox{\boldmath $\xi$}}^\varphi_{\text{m},tr}+2i\text{m}M h_{t\varphi}
-6Mr\sin^2\vartheta{\mbox{\boldmath $\xi$}}^r_{\text{m},t}+2r^2\sin^2\vartheta 
{\mbox{\boldmath $\xi$}}^r_{\text{m},t} \nonumber\\
&-&\sin2\vartheta(3M h_{t\vartheta}-rh_{t\vartheta})
-3Mh_{\varphi\varphi,t}+rh_{\varphi\varphi,t}r \} \\
{{\mbox{\boldmath $\xi$}}^\varphi_{\text{m},\vartheta}}&=&-\frac{i}
{(2\,{r}^{2}\left ({\text{m}}^{2}+1\right )\sin^3\vartheta)}\Bigg(-2\,i
\sin\vartheta\cos^2\vartheta{\it h_{\vartheta\varphi}}+2\,\text{m}\cos\vartheta{
\it h_{\varphi\varphi}}+2\,\text{m}\cos\vartheta\sin^2\vartheta
r{{\mbox{\boldmath $\xi$}}^r_{\text{m}}}\nonumber\\
&-&2\,i{h_{\vartheta\varphi}}\,
\sin^3\vartheta-\text{m}\sin^2\vartheta\cos\vartheta
{h_{\vartheta\vartheta}}-\text{m}\sin\vartheta{h_{\varphi\varphi,\vartheta}}
-2\,\text{m}
\sin^3\vartheta r {{\mbox{\boldmath $\xi$}}^r_{\text{m},\vartheta}} \Bigg),
\nonumber\\
{\mbox{\boldmath $\xi$}}^t_{\text{m} ,t}&=& -
\frac{2Mr^2{{\mbox{\boldmath $\xi$}}^r_{\text{m}}}+h_{tt}}{2r(r-2M)}\nonumber
 \\
{\mbox{\boldmath $\xi$}}^\varphi_{\text{m},t}&=& -
\frac{(r-2M)[rh_{t\varphi,r}+r^3\sin^2\vartheta{\mbox{\boldmath $\xi$}}^%
\varphi_{\text{m},rt} +i\text{m}(r-2M){\mbox{\boldmath $\xi$}}^t_{\text{m},r}]
-2M h_{t\varphi}}{2r^2\sin^2\vartheta(r-3M)}
\end{eqnarray}

${\mbox{\boldmath $\xi$}}^\varphi_{,r}$ and
${\mbox{\boldmath $\xi$}}^t_{,r}$ as given by Eqs.\ (\ref{Xphi}) and
(\ref{Xt}) are well defined in the $a\to0$ limit.

\end{document}